# Wavelength Requirements for Life Detection via Reflected Light Spectroscopy of Rocky Exoplanets

Joshua Krissansen-Totton,[1,3,4] Anna Grace Ulses,[1,3,4] Maxwell K. Frissell,[1,2] Samantha Gilbert-Janizek,[4,5] Amber Young,[6] Jacob Lustig-Yaeger,[3,7] Tyler D. Robinson,[3,8] Stephanie Olson,[9] Eleonora Alei,[6,*] Giada Arney,[6] Celeste Hagee,[6,10] Chester Harman,[11] Natalie Hinkel,[12] Émilie Laflèche,[9] Natasha Latouf,[6] Avi Mandell,[6] Mark M. Moussa,[6] Mary N. Parenteau,[13] Sukrit Ranjan,[8] Blair Russell,[14] Edward W. Schwieterman,[3,15] Clara Sousa-Silva,[16] Armen Tokadjian,[17] and Nicholas Wogan[3,11,*]

## Abstract

Searching for signs of life is a primary goal of the Habitable Worlds Observatory (HWO). However, merely detecting oxygen, methane, or other widely discussed biosignatures is insufficient evidence for a biosphere. In parallel with biosignature detection, exoplanet life detection additionally requires characterization of the broader physicochemical context to evaluate planetary habitability and the plausibility that life could produce a particular biosignature in a given environment. Life detection further requires that we can confidently rule out photochemical or geological phenomena that can mimic life (i.e., "false positives"). Evaluating false-positive scenarios may require different observatory specifications than biosignature detection surveys. Here, we explore the coronagraph requirements for assessing habitability and for cautiously excluding known false-positive (and false-negative) scenarios for oxygen and methane, the two most widely discussed biosignatures for Earth-like exoplanets. We find that broad wavelength coverage ranging from the near ultraviolet (UV; 0.26 μm) and extending into the near infrared (NIR; 1.7 μm) is necessary to contextualize these potential biosignatures with HWO. The short-wavelength cutoff is driven by the need to identify Proterozoic-like biospheres via $O_3$, whereas the long-wavelength cutoff is driven by the need to contextualize $O_2$ and $CH_4$ biosignatures via constraints on carbon-bearing atmospheric species. The ability to obtain spectra with signal-to-noise ratios of 20–40 across this 0.26–1.7 μm range (assuming R = 7 UV, R = 140 VIS, and R = 70 NIR) is also required. While not every Earth-analog biosignature and false positive can be unambiguously identified with these capabilities—and the plausibility and contextual clues of many biosignature false positives remain an area of active research—our minimal spectral recommendations would enable a broad search for Earth-like life assuming such observations are achievable for a statistically meaningful number of

[1] Department of Earth and Space Sciences, University of Washington, Seattle, Washington, USA.
[2] Department of Astronomy and Planetary Science, Northern Arizona University, Flagstaff, Arizona, USA.
[3] NASA NExSS Virtual Planetary Laboratory, University of Washington, Seattle, Washington, USA.
[4] Astrobiology Program, University of Washington, Seattle, Washington, USA.
[5] Department of Astronomy, University of Washington, Seattle, Washington, USA.
[6] NASA Goddard Space Flight Center, Greenbelt, Maryland, USA.
[7] JHU Applied Physics Laboratory, Laurel, Maryland, USA.
[8] Lunar and Planetary Laboratory, University of Arizona, Tucson, Arizona, USA.
[9] Department of Earth, Atmospheric, and Planetary Sciences, Purdue University, West Lafayette, Indiana, USA.
[10] Southeastern Universities Research Association, Washington, District of Columbia, USA.
[11] Planetary Systems Branch, Space Science Division, NASA Ames Research Center, Moffett Field, California, USA.
[12] Department of Physics & Astronomy, Louisiana State University, Baton Rouge, Louisiana, USA.
[13] Exobiology Branch, NASA Ames Research Center, Moffett Field, California, USA.
[14] Schmid College of Science and Technology, Chapman University, Orange, California, USA.
[15] Department of Earth and Planetary Sciences, University of California, Riverside, Riverside, California, USA.
[16] Bard College, Annandale-On-Hudson, New York, USA.
[17] Jet Propulsion Laboratory, California Institute of Technology, Pasadena, California, USA.
*NASA Postdoctoral Program Fellow.

HWO targets. Key Words: Biosignatures—Habitable Worlds Observatory—Retrieval—Biosignature false positives. Astrobiology 00, 000–000.

## 1. Introduction

At the time of writing, no ground- or space-based telescope has the capability to detect life on temperate rocky planets around Sun-like stars. In principle, the James Webb Space Telescope (JWST) has the requisite precision to search for biosignatures around a handful of temperate transiting exoplanets orbiting low mass M dwarf stars (Meadows et al., 2023; Seager et al., 2025), although the success of this search is contingent on overcoming challenges due to stellar contamination (Rackham et al., 2018; de Wit et al., 2024; Rathcke et al., 2025) and diminished absorption signals due to small scale heights and aerosols (Suissa et al., 2020; Fauchez et al., 2022). Given the overwhelming scientific and public interest in the search for life beyond Earth, the 2020 Astrophysics Decadal Survey recommended a $\sim$6 m ultraviolet (UV)/visible (VIS)/near infrared (NIR) space telescope to search for biosignatures on $\sim$25 nearby Earth-sized planets in the habitable zone of Sun-like stars (National Academies of Sciences, Engineering, and Medicine, 2021). This telescope, currently referred to as the Habitable Worlds Observatory (HWO), would succeed the Roman Space Telescope as NASA's next astrophysics flagship. HWO would directly image potentially habitable planets using coronagraphic starlight suppression and obtain reflected light spectra to constrain atmospheric and surface compositions. Spectroscopic characterization would include searching for atmospheric and surface biosignatures such as waste gases produced by life or biological pigments (Des Marais et al., 2002; Seager et al., 2005; Meadows et al., 2018; Schwieterman et al., 2018; Schwieterman and Leung, 2024).

While it is challenging to anticipate the variety and prevalence of biosignatures that may exist elsewhere, our search is guided by universal features of life and inferences about energy metabolisms in plausible planetary environments (Krissansen-Totton et al., 2022). Life requires a liquid solvent; water is the most likely candidate given cosmochemical abundances, its stability over a wide temperature range, and chemical properties conducive to molecular information processing (Pohorille and Pratt, 2012). Carbon-based (organic) chemistry also seems like the only possibility for sustaining diverse biochemical structures (Petkowski et al., 2020). Carbon-based chemistry implies that methane-producing life is likely a common metabolic strategy that may evolve on other worlds given the ubiquity of the necessary substrates ($H_2$, $CO_2$) and the probable inevitability of methane production via degradation of organic matter (Thompson et al., 2022). Indeed, methanogenesis evolved early in Earth's history (Weiss et al., 2016; Wolfe and Fournier, 2018). Similarly, oxygen biosignatures from photosynthesis may also be common since the substrate or electron donor used ($H_2O$) would be widely available. Moreover, any surface life that develops the capacity for oxygenic photosynthesis would possess an immense evolutionary advantage over other metabolisms that depend on geochemical energy whose reductants or electron donors are typically in low supply (Krissansen-Totton et al., 2022). In short, if life is common in the universe, then $CH_4$-producing and $O_2$-producing biospheres may also be widespread.

This focus on $CH_4$ and $O_2$ (and its photochemical byproduct $O_3$) as exoplanet biosignatures is also supported by our knowledge of Earth through time. Figure 1 shows a schematic of Earth's atmospheric evolution from 4 Ga (billion years ago) to the present. Prior to the oxygenation of the atmosphere by cyanobacteria, atmospheric methane was likely abundant and mostly produced by life (Catling et al., 2001; Kasting, 2005; Thompson et al., 2022; Ebadirad et al., 2025). Similarly, virtually all the atmospheric $O_2$ (and $O_3$) present in Earth's atmosphere post $\sim$2.4 Ga is a byproduct of oxygenic photosynthesis (Catling and Claire, 2005; Lyons et al., 2014). The right-hand panel of Figure 1 shows that these biogenic gases ($O_2$, $O_3$, $CH_4$) produced clear features in Earth's reflected light spectrum; atmospheric biosignatures have likely been present in Earth's atmosphere for the last 4 billion years.

With that said, the detection of $O_2$ or $CH_4$ would not constitute life detection since there are a number of scenarios involving nonbiological processes that could potentially produce $O_2$- or $CH_4$-rich atmospheres, the so-called biosignature "false positives" (Domagal-Goldman et al., 2014; Meadows et al., 2018; Thompson et al., 2022). False positives must be carefully considered and ruled out to increase confidence that a potential biosignature is truly due to life (e.g., Meadows et al., 2022). For oxygen, the potential false positives most relevant to HWO targets include $O_2$ build-up due to a lack of a noncondensable background gas leading to elevated $H_2O$ photolysis and H escape (Wordsworth and Pierrehumbert, 2014; Kleinböhl et al., 2018) and waterworld false positives, whereby large surface water inventories suppress geologic oxygen sinks, which leads to gradual $O_2$ accumulation via H escape (Krissansen-Totton et al., 2021). Photochemical false positives due to continuous photodissociation of $CO_2$ into CO and $O_2$ are generally less likely around F/G/K stars (Harman et al., 2018; Ranjan et al., 2020, 2023; Barth et al., 2024), which will constitute the bulk of HWO targets (Mamajek and Stapelfeldt, 2024; Tuchow et al., 2024). The most robust photochemical $O_2$ false positive arises for very hydrogen-poor planets, but, fortuitously, these can be falsified by detecting $\geq$1 ppm total hydrogen (Gao et al., 2015), which can be accomplished by detecting the $H_2O$ already required to judge the planet habitable. Nonetheless, photochemical false positives are still considered here because (1) HWO will be able to characterize a small number of M dwarf–hosted planets, and (2) remaining uncertainties in photochemical processes and boundary conditions mean they cannot be completely ruled out. Methane false positives have been less exhaustively studied, but plausible scenarios include methane-degassing from an extremely reduced planetary interior (Schindler and Kasting, 2000; Wogan et al., 2020; Liggins et al., 2023) or methane production via Fischer–Tropsch-type (FTT) reactions in hydrothermal settings (Fiebig et al., 2007; Etiope and Sherwood Lollar, 2013; Guzmán-Marmolejo et al., 2013; Krissansen-

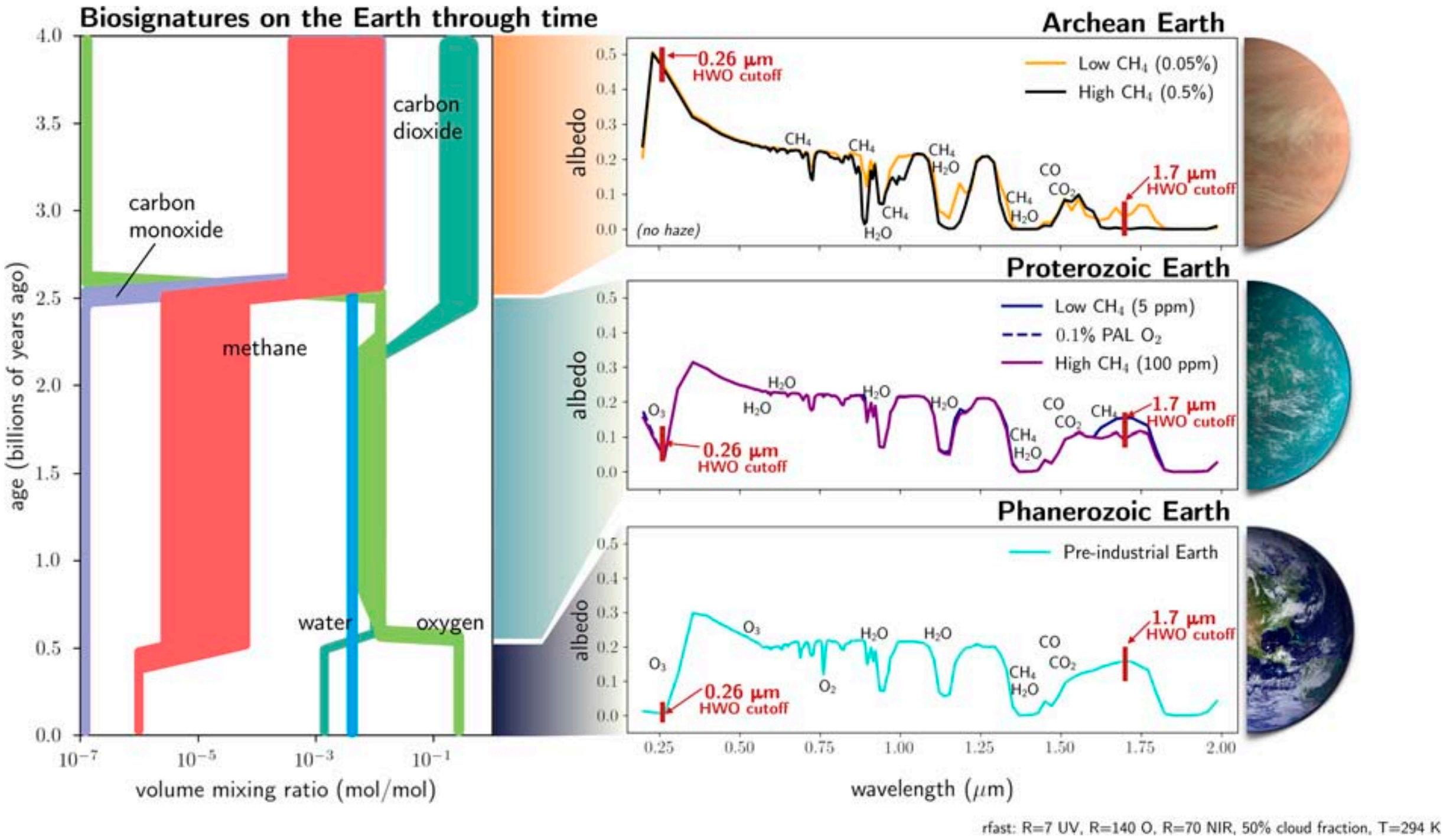

**FIG. 1.** Earth biosignatures through time and their reflected light spectra. The left subplot shows estimates of Earth's bulk atmospheric composition through time (advancing from top to bottom), and the figures on the right show the corresponding reflected light spectra considered in this study. Key absorption features for biosignature gases ($O_3$, $O_2$, $CH_4$) and suggested long- and short-wavelength cutoffs are highlighted (as discussed further in Sections 3 and 4). Adapted from National Academies of Sciences, Engineering, and Medicine (2021) and LUVOIR and HabEx Final Reports (LUVOIR Team, 2019; Gaudi et al., 2020).

Totton et al., 2018b). We assume that biosignature false positives due to $CH_4$ in both sub-Neptune atmospheres (Damiano and Hu, 2021; Madhusudhan et al., 2023; Benneke et al., 2024; Hu et al., 2025a) and Titan analogs (Thompson et al., 2022; Levi et al., 2023), and $O_2$ in runaway greenhouse atmospheres (Kasting and Pollack, 1983) will be identifiable from other contextual clues, and they are not considered further here. For example, rocky planets and sub-Neptunes/waterworlds may be distinguished via Rayleigh scattering slope, atmospheric constituents, or broadband polarimetry (Hu et al., 2025b); Titan analogs would have conspicuous bulk densities; and $O_2$–$O_2$ collisionally induced absorption (CIA) features could identify post-runaway atmospheres (Schwieterman et al., 2016).

The coevolution of life and the environment on Earth (Fig. 1) also raises the possibility of "false negatives," which are productive surface biospheres that are challenging to detect due to instrumental limitations (Reinhard et al., 2017). The canonical example of a false negative is Proterozoic Earth, whereby atmospheric $O_2$ was present but likely undetectable in reflected light for HWO-like architectures; if $CH_4$ were similarly undetectable due to low abundances (Olson et al., 2016; Laakso and Schrag, 2019), then the Proterozoic biosphere would be undetectable unless UV spectroscopy enabled $O_3$ detection via Hartley band absorption (0.2–0.32 μm).

To ensure that HWO can fulfill its intended purpose of searching for life on $\sim 25$ potentially habitable rocky planets, the telescope must be designed to not only detect potential biosignatures but also, ideally, rule out known false positives and false negatives for the target sample. This will necessitate sufficient wavelength coverage, resolution, and precision to constrain the planetary context and look for atmospheric or surface indicators that disfavor the abovementioned scenarios.

### *1.1. Prior explorations of exoplanet biosignature detectability*

Explorations of potential exoplanet biosignature gases long predate the discovery of the first exoplanet (Lovelock, 1975; Owen, 1980). Sagan et al. (1993) used spatially resolved Galileo spacecraft observations to demonstrate the feasibility of detecting present-day Earth life (for instance via 0.76 $\mu$m $O_2$ detection). Des Marais et al. (2002) calculated the size of spectral features in reflected light for a range of gas abundances, and they estimated wavelength ranges required for detecting $CO_2 + CH_4$ in the NIR, $O_2$ in the visible, and potentially $O_3$ at UV wavelengths. Woolf et al. (2002) used Earthshine data to observe present-day Earth as if it were a spatially unresolved exoplanet and demonstrated $O_2 + H_2O$ would be detectable. The challenges and opportunities provided by rotational and orbital variability have also been a focus of numerous (exo-)Earth-observation simulations (Ford et al., 2001; Goode et al., 2001; Robinson et al., 2010). These early explorations of exoplanet characterization informed the design of several mission concepts intended to search for life via reflected light spectroscopy (Traub et al., 2006; Stapelfeldt et al., 2014), as well as simulations of coronagraph optimization (Kasdin et al., 2003) and survey completeness (Brown, 2005; Brown and Soummer, 2010).

More recently, simulated Bayesian retrieval studies have revisited the detectability of Earth-like atmospheric biosignatures and their false positives via reflected light spectroscopy. Feng et al. (2018) performed visible (0.4–1.0 μm) reflected light retrievals to show that $O_2$ and $O_3$ are detectable for modern Earth analogs and that background pressure constraints imply that $N_2$ can be indirectly constrained. Damiano and Hu (2022) authored the first retrieval study to explicitly explore the longwave cutoff: they investigated both modern Earth and Archean Earth analogs with synthetic reflected light retrievals and found that 0.4–1.0 μm is sufficient for $O_2$ detection but that extending wavelength coverage to 1.8 μm is necessary both to constrain the bulk background gas ($O_2$, $N_2$, $CO_2$) and to identify Archean Earth-like $CH_4$ biosignatures. Hall et al. (2023) performed retrievals that show modern Earth ($N_2$–$O_2$) analogs and CO-rich, $O_2$ false positives can be disentangled so long as wavelength coverage extends to at least 1.6 μm to encompass CO absorption features. However, neither study explored the marginal benefit from partial NIR spectra nor did these studies explicitly consider the extent to which different false-positive scenarios for Earth through time could be excluded. Damiano et al. (2023) explored short-wavelength cutoff via synthetic reflected light retrievals and found that a 0.25-μm cutoff is sufficient for Proterozoic $O_3$ detection (and $O_2$ constraints) as it provided similar constraints to a 0.2-μm cutoff, but the cost/benefit of alternative short-wavelength cutoffs was not explicitly modeled. Young et al. (2024b) investigated the detectability of $O_2$ and $CH_4$ biosignatures on Proterozoic Earth but only for the most optimistic (0.2–1.8 μm) total wavelength range. Similarly, Gilbert-Janizek et al. (2024) investigated the ability to detect and interpret biosignature gas pairs for a Modern Earth twin in reflected light with realistic noise and found that $CH_4$ and $CO_2$ are difficult to detect in the NIR even for an optimistic long-wavelength cutoff of 2.0 μm. However, they also showed that $O_3$ in the near-UV (NUV) is readily detectable, and its atmospheric vertical structure may even be inferred via the shape of the Hartley band. Latouf et al. (2025) conducted retrievals with variable amounts of $CH_4$ that represent different times in Earth's history, although they focused on the bandpass width needed for unambiguous $CH_4$ detection rather than sensitivity to total wavelength coverage. The closest previous work to this study is that of Tokadjian et al. (2024), who investigated the detectability of Archean Earth-like $CH_4$ biosignatures (and CO-rich false positives) for a range of long-wavelength cutoffs.

### 1.2. This study in context

What is missing from the existing literature is a systematic exploration of precise wavelength ranges and corresponding signal-to-noise (SNR) needed to detect Earth-like biosignatures throughout Earth history and disfavor their most relevant false positives. While previous retrieval studies have provided some indication of what wavelength ranges are needed, it is not known exactly how much of the UV and NIR spectra are needed to characterize Earth analogs, and the detectability of their false positives is less explored. Exact cutoffs are important because there are considerable engineering challenges associated with pushing coronagraphic capabilities into the UV (Balasubramanian et al., 2011; Pueyo et al., 2019), whereas additional NIR coverage demands more aggressive inner working angle (IWA) constraints and, potentially, an ambitious passive cooling strategy to minimize thermal background noise (LUVOIR Team, 2019). Thus, given that HWO architectures and instrument trades are currently being investigated, a thorough exploration of the minimum wavelength range needed for life detection is timely. Here, we present a suite of synthetic retrievals to determine wavelength range required to characterize Earth-through-time biosignatures and discriminate them from most known false-positive scenarios.

## 2. Methods

### 2.1. Radiative transfer and retrieval model

We used the *rfast* radiative transfer retrieval suite, a one-dimensional exoplanetary atmosphere forward model from Robinson and Salvador (2023). Molecular opacities in *rfast* are derived from the HITRAN2020 database (Gordon et al., 2022) using the Line-By-Line ABsorption Coefficients tool (Meadows and Crisp, 1996). The planetary spectrum resolution is then degraded to match the resolution specified via an instrument model. To optimize the retrieval process, *rfast* is nearly entirely written using linear algebra techniques to take advantage of vectorized computational methods. We also made use of *rfast*'s blended water ice/liquid gray cloud model (Robinson and Salvador, 2023).

After generating noisy exoplanet spectra, *rfast* retrievals were handled via the *emcee* Markov chain Monte Carlo (MCMC) sampler (Foreman-Mackey et al., 2013). *rfast* allows for retrievals of atmospheric, planetary, and orbital parameters with uniform or Gaussian priors in either log or linear space. Here, all 14 retrieved planetary parameters were retrieved in $\log_{10}$ space (Table 1). Our retrievals included seven unknown atmospheric gas partial pressures: $N_2$, $O_2$, $H_2O$, $O_3$, $CO_2$, $CH_4$, and CO. A $\log_{10}$ uniform prior was assumed for all constituents ranging from $10^{-2}$ to $10^7$ Pa except for potentially trace species $O_2$, $O_3$, and CO, where the prior range was extended to $10^{-12}$ to $10^7$ Pa (all gases have a negligible impact on the reflected light spectrum with abundances below the assumed lower prior bound). Unless stated otherwise, our retrievals also included seven unknown planetary parameters: surface albedo (As, prior range 0.01–1.0), planet radius (Rp, prior range $10^{-0.5}$ to $10^{0.5}$ $R_\oplus$), planet mass (Mp, prior range 0.1–10 $M_\oplus$), cloud thickness ($\Delta p_c$, prior range 1–$10^7$ Pa), cloud top pressure ($p_t$, prior range 1–$10^7$ Pa), cloud extinction optical thickness ($\tau_c$, prior range $10^{-3}$ to $10^3$), and cloud cover fraction ($f_c$, prior range $10^{-3}$ to 1). Cloud top pressures below the surface pressure were prohibited, and a gray surface albedo was assumed except for the water-world false-positive cases described below, which follow the methodology of Ulses et al. (2025). Each retrieval adopted 200 "walkers" within *emcee,* and the walker positions were randomly initialized within each parameter's prior ranges. Consistent with previous retrieval studies, our retrievals had each walker take 200,000 steps with a burn-in of 100,000–150,000 and a thinning of 100 for output compression.

Our calculations assume that $H_2$ is not a major atmospheric constituent in any of our scenarios. This is justified for oxidized (Phanerozoic/Proterozoic) atmospheres because combustible $H_2$–$O_2$ atmospheres are kinetically unstable (Grenfell et al., 2018). For more reducing (Archean) terrestrial atmospheres, $H_2$ abundances that can be plausibly replenished by degassing (i.e., <5–10% $H_2$) have a negligible impact on the reflected light spectrum. We additionally assume that $H_2$-rich sub-Neptune atmospheres have already been excluded based

TABLE 1. EARTH-THROUGH-TIME ATMOSPHERIC SCENARIOS AND RETRIEVAL PRIORS

| *Planetary parameter* | *Prior range, $log_{10}$(Pa)* | *Phanerozoic* | *Proterozoic (low $CH_4$)* | *Proterozoic (High $CH_4$)* | *Archean (low $CH_4$)* | *Archean (high $CH_4$)* |
|---|---|---|---|---|---|---|
| $N_2$ | [$10^{-2}$, $10^7$] | 79.6% | 98.5% | 98.8% | 80% | 94.5% |
| $O_2$ | [$10^{-12}$, $10^7$] | 20% | 0.2% | 0.2% | None | None |
| $O_3$ | [$10^{-12}$, $10^7$] | $7 \times 10^{-7}$ | $7 \times 10^{-8}$ | $7 \times 10^{-8}$ | None | None |
| $CO_2$ | [$10^{-2}$, $10^7$] | 0.1% | 1% | 1% | 20% | 5% |
| CO | [$10^{-12}$, $10^7$] | 0.1 ppm | 0.1 ppm | 0.1 ppm | 0.005% | 0.05% |
| $CH_4$ | [$10^{-2}$, $10^7$] | 1 ppm | 5 ppm | 100 ppm | 0.05% | 0.5% |
| $H_2O$ | [$10^{-2}$, $10^7$] | 0.3% | 0.3% | 0.3% | 0.3% | 0.3% |
| Pressure (bar)[a] | N/A[a] | 1 | 1 | 1 | 1 | 1 |
| Surface albedo, As | | | | 0.1 [0.01 to 1.0] | | |
| Radius, Rp ($R_\oplus$) | | | | 1.0 [$10^{-0.5}$ to $10^{0.5}$] | | |
| Mass, Mp ($M_\oplus$) | | | | 1.0 [$10^{-1}$ to $10^1$] | | |
| Cloud thickness, $\Delta p_c$ (Pa) | | | | 10,000 [$10^0$ to $10^7$] | | |
| Cloud top pressure, $p_t$ (Pa) | | | | 60,000 [$10^0$ to $10^7$] | | |
| Cloud extinction, $\tau_c$ | | | | 1.0 [$10^{-3}$ to $10^3$] | | |
| Cloud fraction, $f_c$ | | | | 50% [$10^{-3}$ to $10^0$] | | |

Assumed atmospheric abundances and retrieval prior ranges for the Earth through time calculations, the results from which are shown in Figure 4. Square brackets denote prior range, sampled in $\log_{10}$ space. Prior ranges for all gas species are $10^{-2}$ to $10^7$ Pa (except for potentially trace species), with retrievals performed in partial pressure space.

[a]Pressure is not an independent free parameter as it is the sum of the constituent partial pressures (partial pressures are retrieved parameters).

on contextual information, as described above, although we acknowledge that this may be challenging in practice.

For the majority of retrievals in this study, we opted to perform partial pressure retrievals to avoid biased priors compared with retrievals on mixing ratios. A partial pressure retrieval is broadly equivalent to retrieving centered log ratios for atmospheric abundances (Benneke and Seager, 2012; Damiano and Hu, 2021, 2022), except that it removes the need to retrieve total pressure as a separate parameter ($p_{max}$ is simply the sum of the partial pressures of every gas in the atmosphere) and avoids any potential biases with MCMC chain initialization in centered-log-ratio space. Adopting uniform partial pressure priors may create bias toward low total pressures, but we do not see any evidence of this in retrieved posteriors. Mixing ratio retrievals were used for Archean cases that otherwise were poorly converged; this should not impact Archean conclusion results since total pressure and background gas constraints are of secondary importance for interpreting Archean-analog scenarios.

Nominal retrievals were performed with spectral resolutions, $\lambda/\Delta\lambda$, of 7, 140, and 70 corresponding to the UV (0.2–0.4 $\mu$m), optical (0.4–1.0 $\mu$m), and NIR (1.0–2.0 $\mu$m) portions of the spectrum, respectively. The total wavelength range of each retrieval is an independent variable that we systematically investigated. An SNR of 20 was applied at 1.0 $\mu$m with constant error bars and then applied across the full wavelength range. SNR = 20 was assumed in all nominal retrievals, but sensitivity tests for SNR = 10 and SNR = 40 were also performed. An isothermal Earth-like temperature was assumed, and previous work has shown that reflected light retrievals are largely insensitive to temperature within the range expected for Earth analogs (Hall et al., 2023; Gilbert-Janizek et al., 2024).

Using our retrieval outputs from *rfast* and *emcee*, we determined what SNR and resolution would be necessary to constrain atmospheric constituents sufficiently well to rule out or render unlikely known false positives. For a more in-depth analysis of *rfast*'s capabilities, see Robinson and Salvador (2023).

### 2.2. Earth-through-time scenarios

Table 1 summarizes the representative Earth-through-time atmospheric abundances explored in this study. For Phanerozoic Earth (0.54–0 Ga), we used modern values for $pN_2$, $pO_2$, $pCH_4$, and pCO, but slightly higher $pCO_2$ values (0.1%), which were more representative of the long-term average (Foster et al., 2017; Witkowski et al., 2018; Steinthorsdottir et al., 2025). For Proterozoic Earth (2.5–0.54 Ga), we considered two endmember cases, one with $CH_4$ at the high end of model estimates, around 100 ppm (Kasting, 2005; Zhao et al., 2018), and another with $CH_4$ at the low end of model estimates, around 5 ppm (Olson et al., 2016; Laakso and Schrag, 2019). Proterozoic oxygen/ozone levels were assumed to be around 1% of the present atmospheric level (PAL) in both cases (Lyons et al., 2021; Wogan et al., 2022), although we considered lower $pO_2$ (0.1% PAL) when investigating $O_3$ detectability and the short-wavelength cutoff (see Supplementary Data). Atmospheric $CO_2$ was kept at 1% to broadly represent the Proterozoic average (Krissansen-Totton et al., 2018a). Similarly, we considered two Archean cases broadly representative of the low $CH_4$ (0.05%) early Archean (4.0–3.2 Ga), and the high $CH_4$ (0.5%) late Archean (3.2–2.5 Ga) (Zahnle et al., 2019; Catling and Zahnle, 2020). Archean $pCO_2$ values were also chosen to reflect this evolution, 20% and 5% for the early and late Archean, respectively (Catling and Zahnle, 2020). Based on ecosystem models, we chose corresponding CO values representative of plausible $CH_4$:CO ratios, with 0.005% and 0.05% for the early and late Archean, respectively (Schwieterman et al., 2019; Sauterey et al., 2020). Atmospheric water vapor was assumed to be 0.3%, which is the well-mixed equivalent of the modern climate. All cases assumed a 1 bar total atmospheric pressure, which is implicitly retrieved via the sum of the free gas partial pressure parameters.

Note that retrieval setup and Earth-through-time abundances were slightly modified for shortwave cutoff $O_3$-detectability retrievals, as described in Supplementary Data.

Table 2. Biosignature False Positives and Their Contextual Discriminants

| *Oxic or anoxic atmospheres* | *False-positive scenario* | *Contextual information needed to render false positive unlikely* |
|---|---|---|
| False positives for oxygen/ozone biosignatures (Phanerozoic and Proterozoic Earth) | **Low noncondensable inventory** (Wordsworth and Pierrehumbert, 2014). Lack of background $N_2$/$CO_2$/CO results in high $H_2O$ abundance, and rapid $O_2$ accumulation via H escape | Infer background gas by elimination and total pressure Rayleigh slope + pressure broadening (Hall et al., 2023; Young et al., 2024b). Coexisting $CH_4$ in disequilibrium (Krissansen-Totton et al., 2016). |
| | **Waterworld scenario** (Krissansen-Totton et al., 2021). Large surface water inventory suppresses crustal production and oxygen sinks, resulting in $O_2$ accumulation via H escape. | Land detection via surface spectrum (Ulses et al., 2025) or glint and time-resolved mapping (Lustig-Yaeger et al., 2018). Coexisting $CH_4$ in disequilibrium (Krissansen-Totton et al., 2016). |
| | Runaway greenhouse atmosphere (Kasting and Pollack, 1983; Krissansen-Totton et al., 2021). | $CO_2$ and $H_2O$ abundances, total pressure via Rayleigh scattering and $O_2$–$O_2$ CIA (Schwieterman et al., 2016). Habitability indicators. |
| | **Photochemical false positive** (Gao et al., 2015; Harman et al., 2018; Ranjan et al., 2023). Continuous $CO_2$ photodissociation to generate $O_2$ and CO. | *A priori* less likely for F/G/K dwarfs. Stellar UV, $CO_2$, CO, $H_2O$, and NOx abundances to confirm. |
| False positives for methane biosignatures (Archean Earth) | **Reduced mantle degassing** (Wogan et al., 2020; Liggins et al., 2023). Magmatic degassing for $CH_4$ from ultra reducing planetary interior. | Well-constrained abundances of $CH_4$, CO, and $CO_2$ (Thompson et al., 2022). |
| | Warm Titan-analog (Krissansen-Totton et al., 2022). | Bulk density, orbital configuration + evolutionary models. |
| | **Hydrothermally generated methane** (Thompson et al., 2022). | Well-constrained $CH_4$ abundances and stellar UV to infer $CH_4$ fluxes. |
| | C-rich sub-Neptune atmosphere (Benneke et al., 2024). | Constrain mean molecular weight, CO abundance. Habitability indicators. |

Known biosignature false-positive scenarios and their contextual clues for rocky planets around F/G/K stars. Only bolded scenarios with shaded backgrounds are considered in this study.
CIA = collisionally induced absorption.

### 2.3. Biosignature false-positive scenarios

Table 2 summarizes the known false-positive scenarios for oxygen and methane biosignatures alongside the contextual information needed to discriminate them from genuine biosignatures. For this study, we considered only low noncondensable oxygen false positives (Wordsworth and Pierrehumbert, 2014), waterworld oxygen false positives (Krissansen-Totton et al., 2021), photochemical oxygen false positives (Gao et al., 2015), methane degassing false positives (Wogan et al., 2020), and methane production via FTT reactions in hydrothermal settings (Thompson et al., 2022). We also ignored $O_2$ false positives due to H escape

Table 3. Abundances for False-Positive Scenarios

| | *False positives for oxygen/ozone biosignatures (Phanerozoic and Proterozoic Earth)* | | | *False positives for methane biosignatures (Archean Earth)* | |
|---|---|---|---|---|---|
| *Atmospheric contituent* | *Low noncondensable false positive* | *Waterworld false positive* | *Photochemical oxygen false positive* | *Reduced mantle degassing* | *Hydrothermally generated methane* |
| $N_2$ | 0.01 ppm | 79.9% | None | 88.9% | 99% |
| $O_2$ | 99.5% | 20% | 17% | None | None |
| $O_3$ | $7 \times 10^{-7}$ | $7 \times 10^{-7}$ | $10^{-6}$ | None | None |
| $CO_2$ | None | 400 ppm | 46% | 10% | 0.3% |
| CO | 0.01 ppm | 0.1 ppm | 36% | 1% | $2.7 \times 10^{-7}$ |
| $CH_4$ | None | None | None | 0.1% | $1.6 \times 10^{-6}$ |
| $H_2O$ | 0.5% | 0.3% | 0.1 ppm | 0.3% | 0.63% |
| Pressure (bar)[a] | 0.2 | 1 | 1 | 1 | 1 |

Assumed abundances for biosignature false-positive scenarios, the results from which are shown in Figure 5.
[a]Pressure is not an independent free parameter as it is the sum of the constituent partial pressures (partial pressures are retrieved parameters).

during the pre-main sequence (Luger and Barnes, 2015), since this scenario is most relevant to M dwarfs, which do not constitute the majority of HWO targets.

Table 3 shows the assumed atmospheric abundances for the five false-positive scenarios considered here; all other parameters and priors are identical to those in Table 1. Atmospheric abundances are broadly representative of previous studies of these false positives (Wordsworth and Pierrehumbert, 2014; Gao et al., 2015; Wogan et al., 2020; Krissansen-Totton et al., 2021; Thompson et al., 2022), although future work ought to more thoroughly explore the likely parameter space of these biosignature false positives.

### *2.4. Criteria for assessing detectability of atmospheric constituents*

Given the large number of simulated retrievals in this study, we require a quantitative metric for assessing whether individual parameters (like gas mixing ratios) are constrained. Here, we loosely followed the approach of Konrad et al. (2022), whereby we fitted gas mixing ratio 1D marginalized posteriors with analytic functions that represent nondetections (flat or upward sloping functions), detections (Gaussian), or upper limits (a more complex analytic function that allows the posterior to be unbounded below and drop to zero at values less than 1). Posteriors were categorized as nondetections, detections, or upper limits based on which analytic function provides the best fit, and quantitative uncertainties are extracted from the best-fit functions. Note that, for upper limits, the reported $1\sigma$ uncertainty is the uncertainty in the upper bound of the fitted analytic function, not the credible percentile interval, since the latter depends on the choice of prior range. Details of this approach along with illustrative examples are described in Supplementary Data.

## 3. Results

### *3.1. Long-wavelength cutoff*

First, we examine the long-wavelength cutoff needed to confidently identify oxygen and methane biosignatures and render unlikely known nonbiological false positives. Figure 2 shows an example retrieval for early Archean Earth (SNR = 20) using an illustrative wavelength range (0.25–1.8 μm). Figure 2 shows the true and retrieved reflected spectra (inset) alongside the corner plot of retrieved posteriors for all 14 unknown parameters in Table 1. We find that we can readily constrain biogenic $CH_4$ sufficiently well to disfavor the hydrothermal methane false positive (if observations can be obtained at sufficiently high SNR). This is because the retrieved methane abundance is larger than that expected from hydrothermal and FTT reactions (Table 3). The reducing outgassing scenario for abiotic $CH_4$ is more challenging to confidently exclude since the constraints on $CH_4$:CO ratio are crude, but we can at least rule out the CO-dominated atmospheres that might be expected for highly reduced interiors (see Section 5). Next, we consider how the ability to exclude false positives varies for different long-wavelength cutoffs at different times in Earth's history.

3.1.1. Identifying Earth-through-time biosignatures. Figure 3 compares gas abundance posteriors for early Archean retrievals with different long-wavelength cutoffs (SNR = 20). The total wavelength ranges considered include 0.2–X μm where $X$ = 1.0, 1.5, 1.6, 1.7, 1.8, or 2.0 μm (indicated by colored lines and shaded regions). Biogenic $CH_4$ is detectable no matter the long-wavelength cutoff because of prominent features between 0.8 and 1.0 μm (Figs. 1 and 2). However, without a spectrum extending to at least 1.5 μm, there would be no constraints on $CO_2$, which is a key habitability indicator. Moreover, a long-wavelength cutoff of at least 1.6 μm is needed to rule out CO as a bulk constituent, as might be expected for methane produced by magmatic degassing from a reduced interior (Wogan et al., 2020; Liggins et al., 2023). For this early Archean case, the benefits from extending wavelength coverage beyond 1.6 μm are marginal for all gases considered here.

Figure 4 summarizes biosignature interpretation results of spectral retrieval analyses for the long-wavelength coronagraph cutoff. Specifically, we show selected gas constraints for Phanerozoic, Proterozoic (high and low $CH_4$), and Archean scenarios (high and low $CH_4$) for a range of long-wavelength cutoffs and assumed SNRs. Our goal is to systematically investigate the long-wavelength cutoff and SNRs required to contextualize Earth-through-time biosignatures and disfavor known false positives. The color scheme in Figure 4 denotes whether each gas can be detected (green), whether useful upper limits can be obtained (yellow), or whether the gas abundance is unconstrained or the upper limit is uninformative for the purposes of biosignature interpretation (red). For example, CO upper limits are either yellow (CO upper limit < 10%) or red (CO upper limit ≥10%) depending on whether a CO-rich bulk atmosphere can be confidently excluded.

For Phanerozoic Earth (0.54–0 Ga), we find that detecting atmospheric $CO_2$, a potential habitability indicator, would be challenging for any plausible long-wavelength cutoff at the spectral resolutions and SNRs considered here, as would detecting trace amounts of methane. However, access to >1.6 $\mu$m would enable CO-dominated atmospheres to be excluded, thereby ruling out photochemical oxygen false positives. Moreover, an average SNR of at least 20—and ideally 40—is needed to rule out CO as a bulk constituent (for lower SNR = 10, access to ≥1.7 $\mu$m is needed to obtain a CO upper bound).

For Proterozoic Earth (2.5–0.54 Ga), biogenic methane may be detectable if abundances were at the high end of literature estimates ($\sim$100 ppm), and access to ≥1.7 $\mu$m enables biogenic methane detection even if the achievable SNR in the NIR is relatively modest ($\sim$10). This is particularly desirable given the opportunity to simultaneously detect $O_3$ and $CH_4$ under this high $CH_4$ scenario, which would be an especially compelling biosignature with no known false positives. Moreover, access to ≥1.6 $\mu$m with SNR ≥20 enables CO-rich atmospheres to be excluded (or ≥1.7 $\mu$m for SNR ≥10), once again ruling out photochemical oxygen false positives.

Finally, for Archean Earth (4.0–2.5 Ga), both $CO_2$ and biogenic $CH_4$ are detectable for all long-wavelength cutoffs we considered due to their higher assumed abundances. However, access to >1.6 $\mu$m is needed to rule out CO-dominated atmospheres and thereby disfavor methane biosignature false positives produced by a reducing planetary interior; as with Figure 3 above, the Archean CO upper limits are quite crude and would not necessarily fully exclude a reduced degassing scenario. As with the Phanerozoic and Proterozoic cases, SNR ≥20 is necessary to rule out high CO bulk abundances. In fact, CO upper limits are typically poorer for early Archean cases because the $CO_2$ mixing ratio is high (20%) and CO–$CO_2$ absorption features overlap around 1.5–1.6 $\mu$m.

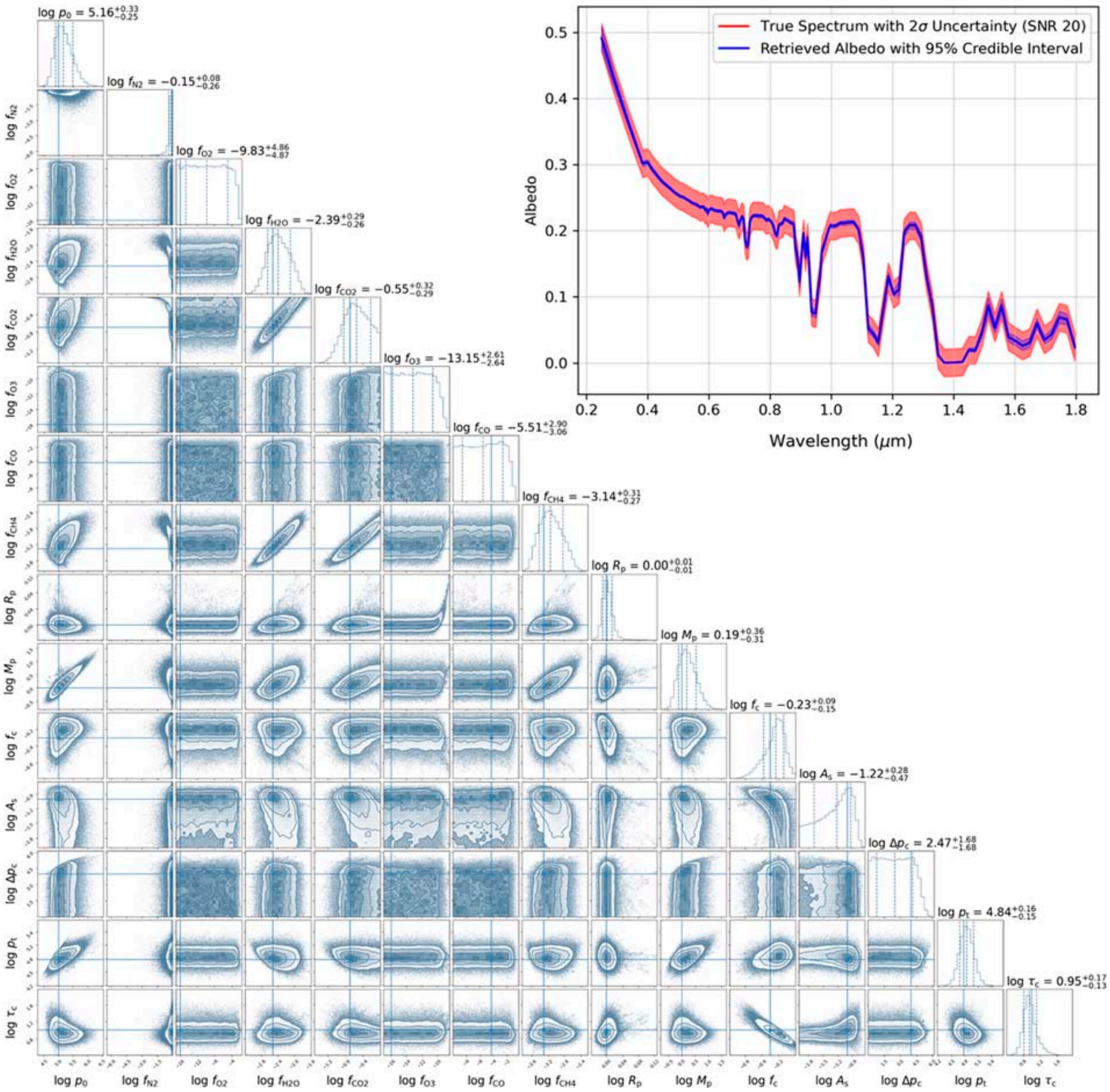

**FIG. 2.** Example atmospheric retrieval for the early Archean scenario using *rfast*. **Corner Plot:** The 2D marginalized posterior distributions are shown for all pairs of parameters. Diagonal distributions show the 1D marginalized posteriors for each parameter, and we list the median retrieved value with uncertainties that indicate the 68% credible interval. Dashed lines (left to right) mark the 16%, 50%, and 84% quantiles. Vertical and horizontal solid blue lines denote true values. Table 1 summarizes the model parameters of this corner plot in the order they are plotted here. **Inset:** Reflectance spectrum of "true" early Archean Earth with assumed SNR = 20, $2\sigma$ uncertainty (R = 7 in UV; R = 140 in VIS; R = 70 in NIR) in red overplotted with the median retrieved spectrum in blue with shaded 95% credible interval. NIR = near infrared; SNR = signal-to-noise; UV = ultraviolet; VIS = visible light.

Based on these results, we find a long-wavelength cutoff of at least 1.7 $\mu$m with NIR SNR ≥20 is needed to constrain $CO_2$, $CH_4$, and CO sufficiently well to detect life on Earth throughout its history (assuming adequate short-wave coverage, as described below). A cutoff of 1.6 $\mu$m with SNR ≥20 may be acceptable if such high SNRs can be reliably achieved at the edge of the NIR spectrum for every target (see Section 5), but a 1.7-$\mu$m cutoff more reliably provides CO constraints and $CH_4$ detections, even with a lower SNR = 10. An even more restricted wavelength range would potentially lead to ambiguous findings as known biosignature false positives could not be excluded.

3.1.2. Identifying biosignature false-positive scenarios. Next, we repeat the analysis described above except considering the false-positive scenarios summarized in Table 3. If life is uncommon but biosignature false positives are common, then it is important that HWO have the capabilities to identify such false-positive scenarios to obtain useful constraints on the prevalence of life.

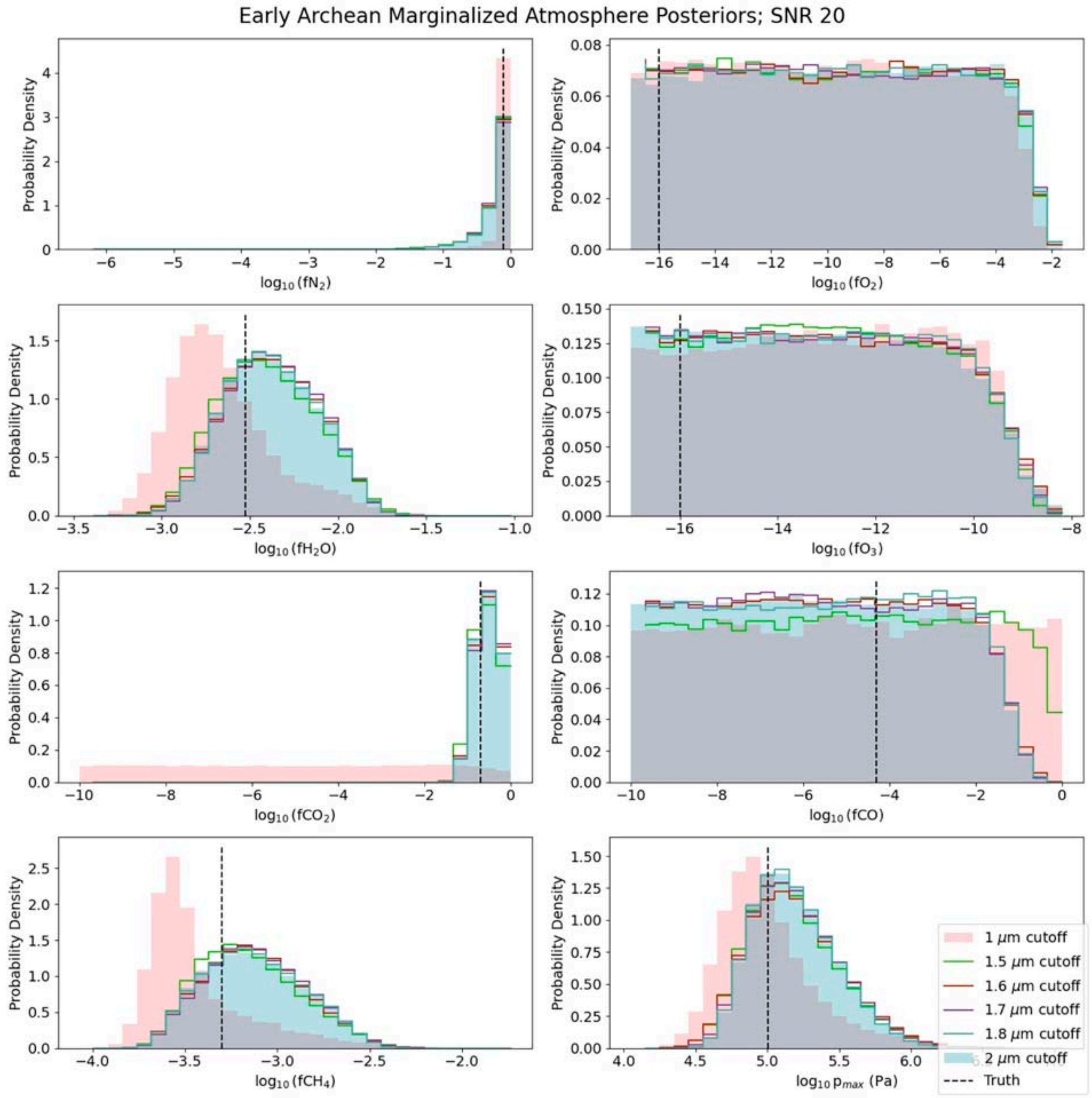

**FIG. 3.** Atmospheric mixing ratios and total pressure marginalized posteriors for the early Archean SNR = 20 scenario. Each subplot shows the posterior probability density for the labeled parameter, and lines/shadings represent retrievals with different X longwave cutoffs (with a spectrum 0.2–X μm). Vertical dashed lines denote true values. We find that the long-wavelength cutoff of 1.5 μm is needed to detect Archean Earth-like $CO_2$, whereas a cutoff of 1.6 μm is needed to put an upper limit on CO. Other constituents are largely unaffected by the longwave cutoff.

Figure 5 summarizes false-positive results of spectral retrieval analyses for the long-wavelength coronagraph cutoff. Specifically, we show selected constraints for photochemical false positives, waterworld false positives, low noncondensable false positives, reduced mantle methane outgassing, and hydrothermal methane false positives for a range of long-wavelength cutoffs and assumed SNRs. Our goal is to systematically investigate the long-wavelength cutoff and SNRs required to identify such false positives and thereby disfavor a biological interpretation. As above, the color scheme in Figure 5 denotes whether each gas or relevant atmospheric/surface parameter can be detected (green), whether useful upper limits can be obtained (yellow), or whether the gas abundance is unconstrained or the upper limit is uninformative for the purposes of biosignature interpretation (red).

First, we consider oxygen false-positive scenarios. We find that wavelength coverage to at least 1.6 μm, ideally with SNR 40, is needed to confidently constrain CO for a photochemical oxygen false positive. Higher SNR is required because CO features are comparatively weak in $CO_2$-rich atmospheres given their adjacent absorption features around 1.5–1.6 μm. With that said, wavelength coverage to merely 1.5 μm with any SNR would enable robust upper limits on $H_2O$ ($<10^{-5}$), which would help distinguish a photochemical

| Long wavelength cutoff (μm): | 1.5 | | | 1.6 | | | 1.7 | | | 1.8 | | | 2.0 | | |
|---|---|---|---|---|---|---|---|---|---|---|---|---|---|---|---|
| SNR | 10 | 20 | 40 | 10 | 20 | 40 | 10 | 20 | 40 | 10 | 20 | 40 | 10 | 20 | 40 |
| **Phanerozoic Earth** (0-0.54 Ga) | Insufficient to contextualize $O_2/O_3$ biosig | | | Marginal for contextualizing $O_2/O_3$ biosig | | | Sufficient to contextualize $O_2/O_3$ biosig | | | Sufficient to contextualize $O_2/O_3$ biosig | | | Sufficient to contextualize $O_2/O_3$ biosig | | |
| $CO_2$ (0.1%) | $-1.3^{+0.4}_{-0.4}$ UL | $-2.2^{+0.5}_{-0.3}$ UL | $-2.4^{+0.1}_{-0.1}$ UL | $-1.6^{+0.8}_{-0.6}$ UL | $-2.3^{+0.3}_{-0.3}$ UL | $-2.6^{+0.2}_{-0.2}$ UL | $-1.9^{+0.5}_{-0.5}$ UL | $-2.4^{+0.4}_{-0.3}$ UL | $-2.6^{+0.2}_{-0.2}$ UL* | $-1.9^{+0.5}_{-0.4}$ UL | $-2.5^{+0.3}_{-0.3}$ UL | $-2.7^{+0.2}_{-0.2}$ UL | $-1.7^{+0.3}_{-0.3}$ UL | $-2.4^{+0.4}_{-0.3}$ UL | $-3.0^{+0.3}_{-0.4}$ DTC |
| $CH_4$ (1 ppm) | $-4.0^{+0.6}_{-0.9}$ UL | $-4.4^{+0.6}_{-2.7}$ UL | $-4.9^{+0.5}_{-0.4}$ UL | $-3.7^{+0.5}_{-0.5}$ UL | $-4.4^{+0.6}_{-0.6}$ UL | $-4.7^{+0.3}_{-0.4}$ UL | $-4.5^{+0.7}_{-1.0}$ UL | $-4.8^{+0.4}_{-0.4}$ UL | $-5.2^{+0.3}_{-0.4}$ UL | $-4.3^{+0.4}_{-0.6}$ UL | $-4.9^{+0.4}_{-0.3}$ UL | $-5.2^{+0.4}_{-0.3}$ UL | $-4.2^{+0.8}_{-0.8}$ UL | $-4.8^{+0.4}_{-0.3}$ UL | $-5.0^{+0.3}_{-0.3}$ UL |
| CO (0.1 ppm) | NC | NC | NC | NC | $-1.6^{+0.5}_{-0.5}$ UL | $-2.1^{+0.3}_{-0.2}$ UL | $-1.1^{+0.6}_{-0.6}$ UL | $-1.7^{+0.5}_{-0.7}$ UL | $-2.2^{+0.3}_{-0.3}$ UL | $-1.1^{+0.5}_{-0.5}$ UL | $-1.8^{+0.5}_{-0.5}$ UL | $-2.1^{+0.2}_{-0.5}$ UL | $-1.0^{+0.9}_{-0.9}$ UL | $-1.8^{+0.5}_{-0.5}$ UL | $-2.3^{+0.4}_{-0.4}$ UL |
| **Low $CH_4$ Proterozoic Earth** (0.54-2.5 Ga) | Insufficient to contextualize $O_3$ biosig | | | Sufficient to contextualize $O_3$ biosig | | | Sufficient to contextualize $O_3$ biosig | | | Sufficient to contextualize $O_3$ biosig | | | Sufficient to contextualize $O_3$ biosig | | |
| $CO_2$ (1%) | $-1.2^{+0.9}_{-0.4}$ UL | $-1.9^{+0.3}_{-0.3}$ DTC | $-2.0^{+0.2}_{-0.2}$ DTC | $-1.3^{+1.3}_{-0.3}$ UL | $-1.9^{+0.3}_{-0.3}$ DTC | $-2.0^{+0.2}_{-0.2}$ DTC | $-1.3^{+0.6}_{-0.3}$ UL | $-1.9^{+0.3}_{-0.3}$ DTC | $-2.0^{+0.1}_{-0.1}$ DTC | $-1.3^{+0.5}_{-0.4}$ UL | $-2.0^{+0.3}_{-0.3}$ DTC | $-2.0^{+0.1}_{-0.1}$ DTC | $-1.3^{+0.4}_{-0.3}$ UL | $-2.0^{+0.3}_{-0.3}$ DTC | $-2.0^{+0.1}_{-0.1}$ DTC |
| $CH_4$ (5 ppm) | $-4.1^{+0.6}_{-1.1}$ UL | $-4.3^{+0.5}_{-0.3}$ UL | $-4.6^{+0.3}_{-0.3}$ UL | $-3.8^{+0.5}_{-0.6}$ UL | $-4.2^{+0.3}_{-0.3}$ UL | $-4.6^{+0.3}_{-0.1}$ UL | $-4.4^{+0.6}_{-0.6}$ UL | $-4.5^{+0.3}_{-0.3}$ UL | $-4.8^{+0.3}_{-0.1}$ UL | $-4.3^{+0.6}_{-0.6}$ UL | $-4.5^{+0.3}_{-0.3}$ UL | $-4.8^{+0.1}_{-0.1}$ UL | $-4.3^{+0.5}_{-0.5}$ UL | $-4.6^{+0.4}_{-0.1}$ UL | $-4.9^{+0.3}_{-0.1}$ UL |
| CO (0.1 ppm) | NC | NC | $-0.7^{+0.2}_{-0.0}$ UL | NC | $-1.7^{+0.5}_{-0.5}$ UL | $-2.3^{+0.4}_{-0.4}$ UL | $-1.2^{+0.7}_{-0.7}$ UL | $-2.0^{+0.5}_{-0.5}$ UL | $-2.3^{+0.3}_{-0.5}$ UL | $-1.2^{+0.7}_{-0.7}$ UL | $-1.9^{+0.5}_{-0.7}$ UL | $-2.3^{+0.5}_{-0.5}$ UL | $-1.4^{+0.8}_{-0.7}$ UL | $-1.8^{+0.4}_{-1.0}$ UL | $-2.4^{+0.5}_{-0.5}$ UL |
| **High $CH_4$ Proterozoic Earth** (0.54-2.5 Ga) | Insufficient to contextualize $O_3/CH_4$ biosig | | | Insufficient to contextualize $O_3/CH_4$ biosig | | | Sufficient to contextualize $O_3/CH_4$ biosig | | | Sufficient to contextualize $O_3/CH_4$ biosig | | | Sufficient to contextualize $O_3/CH_4$ biosig | | |
| $CO_2$ (1%) | NC | $-1.9^{+0.4}_{-0.3}$ DTC | $-2.0^{+0.2}_{-0.2}$ DTC | NC | $-1.9^{+0.3}_{-0.3}$ DTC | $-2.0^{+0.2}_{-0.2}$ DTC | NC | $-1.9^{+0.4}_{-0.3}$ DTC | $-2.0^{+0.2}_{-0.2}$ DTC | NC | $-1.9^{+0.4}_{-0.3}$ DTC | $-2.0^{+0.2}_{-0.2}$ DTC | NC | $-1.9^{+0.4}_{-0.3}$ DTC | $-2.0^{+0.2}_{-0.1}$ DTC |
| $CH_4$ (100 ppm) | $-3.3^{+0.8}_{-0.5}$ UL | $-3.9^{+0.4}_{-0.3}$ DTC | $-4.0^{+0.2}_{-0.2}$ DTC | $-3.2^{+0.8}_{-0.5}$ UL | $-3.9^{+0.3}_{-0.3}$ DTC | $-4.0^{+0.2}_{-0.2}$ DTC | $-3.8^{+0.6}_{-0.5}$ DTC | $-3.9^{+0.3}_{-0.3}$ DTC | $-4.0^{+0.2}_{-0.1}$ DTC | $-3.8^{+0.6}_{-0.5}$ DTC | $-3.9^{+0.4}_{-0.3}$ DTC | $-4.0^{+0.1}_{-0.1}$ DTC | $-3.8^{+0.5}_{-0.4}$ DTC | $-3.9^{+0.4}_{-0.3}$ DTC | $-4.0^{+0.1}_{-0.1}$ DTC |
| CO (0.1 ppm) | NC | NC | $-1.0^{+0.2}_{-0.2}$ UL | NC | $-1.7^{+0.5}_{-0.5}$ UL | $-2.2^{+0.5}_{-0.8}$ UL | $-1.1^{+0.9}_{-0.9}$ UL | $-1.7^{+0.5}_{-0.5}$ UL | $-2.2^{+0.3}_{-0.6}$ UL | $-1.2^{+0.6}_{-0.9}$ UL | $-1.8^{+0.5}_{-0.5}$ UL | $-2.2^{+0.3}_{-0.5}$ UL | $-1.3^{+0.8}_{-0.9}$ UL | $-1.8^{+0.5}_{-0.7}$ UL | $-2.1^{+0.3}_{-0.5}$ UL |
| **Late Archean Earth (high $CH_4$)** (3.2-2.5 Ga) | Insufficient to contextualize $CH_4$ biosig | | | Marginal for contextualizing $CH_4$ biosig | | | Sufficient to contextualize $CH_4$ biosig | | | Sufficient to contextualize $CH_4$ biosig | | | Sufficient to contextualize $CH_4$ biosig | | |
| $CO_2$ (5%) | $-1.1^{+0.5}_{-0.4}$ DTC | $-1.2^{+0.4}_{-0.3}$ DTC | $-1.2^{+0.2}_{-0.1}$ DTC | $-1.0^{+0.5}_{-0.4}$ DTC | $-1.2^{+0.3}_{-0.2}$ DTC | $-1.3^{+0.1}_{-0.1}$ DTC | $-1.0^{+0.5}_{-0.4}$ DTC | $-1.2^{+0.3}_{-0.2}$ DTC | $-1.3^{+0.1}_{-0.2}$ DTC | $-1.0^{+0.5}_{-0.4}$ DTC | $-1.2^{+0.3}_{-0.2}$ DTC | $-1.3^{+0.1}_{-0.2}$ DTC | $-1.0^{+0.5}_{-0.4}$ DTC | $-1.2^{+0.3}_{-0.2}$ DTC | $-1.3^{+0.1}_{-0.1}$ DTC |
| $CH_4$ (0.5%) | $-2.2^{+0.5}_{-0.2}$ DTC | $-2.2^{+0.4}_{-0.2}$ DTC | $-2.2^{+0.2}_{-0.1}$ DTC | $-2.0^{+0.4}_{-0.3}$ DTC | $-2.2^{+0.3}_{-0.2}$ DTC | $-2.3^{+0.1}_{-0.1}$ DTC | $-2.1^{+0.5}_{-0.3}$ DTC | $-2.2^{+0.3}_{-0.2}$ DTC | $-2.3^{+0.1}_{-0.2}$ DTC | $-2.1^{+0.5}_{-0.3}$ DTC | $-2.2^{+0.3}_{-0.2}$ DTC | $-2.3^{+0.1}_{-0.2}$ DTC | $-2.1^{+0.5}_{-0.3}$ DTC | $-2.2^{+0.3}_{-0.2}$ DTC | $-2.3^{+0.1}_{-0.1}$ DTC |
| CO (500 ppm) | NC | NC | NC | $-1.0^{+0.5}_{-0.6}$ UL | $-1.7^{+0.5}_{-0.4}$ UL | $-2.1^{+0.4}_{-0.4}$ UL | $-1.1^{+0.6}_{-0.5}$ UL | $-1.7^{+0.4}_{-0.5}$ UL | $-2.0^{+0.3}_{-0.4}$ UL | $-1.1^{+0.6}_{-0.5}$ UL | $-1.6^{+0.4}_{-0.4}$ UL | $-2.0^{+0.3}_{-0.4}$ UL | $-1.1^{+0.6}_{-0.5}$ UL | $-1.7^{+0.5}_{-0.6}$ UL | $-2.1^{+0.4}_{-0.4}$ UL |
| **Early Archean Earth (low $CH_4$)** (4.0-3.2 Ga) | Insufficient to contextualize $CH_4$ biosig | | | Sufficient to contextualize $CH_4$ biosig | | | Sufficient to contextualize $CH_4$ biosig | | | Sufficient to contextualize $CH_4$ biosig | | | Sufficient to contextualize $CH_4$ biosig | | |
| $CO_2$ (20%) | $-0.6.^{+0.4}_{-0.4}$ DTC | $-0.6^{+0.3}_{-0.3}$ DTC | $-0.6^{+0.3}_{-0.2}$ DTC | $-0.6^{+0.4}_{-0.4}$ DTC | $-0.5^{+0.3}_{-0.3}$ DTC | $-0.6^{+0.3}_{-0.2}$ DTC | $-0.6^{+0.3}_{-0.4}$ DTC | $-0.5^{+0.3}_{-0.3}$ DTC | $-0.6^{+0.3}_{-0.2}$ DTC | $-0.5^{+0.3}_{-0.4}$ DTC | $-0.5^{+0.3}_{-0.3}$ DTC | $-0.6^{+0.2}_{-0.2}$ DTC | $-0.5^{+0.3}_{-0.4}$ DTC | $-0.5^{+0.3}_{-0.3}$ DTC | $-0.6^{+0.2}_{-0.2}$ DTC |
| $CH_4$ (500 ppm) | $-3.4^{+0.3}_{-0.3}$ DTC | $-3.2^{+0.3}_{-0.3}$ DTC | $-3.2^{+0.3}_{-0.2}$ DTC | $-3.1^{+0.3}_{-0.3}$ DTC | $-3.1^{+0.3}_{-0.3}$ DTC | $-3.2^{+0.2}_{-0.2}$ DTC | $-3.1^{+0.3}_{-0.3}$ DTC | $-3.1^{+0.3}_{-0.3}$ DTC | $-3.2^{+0.2}_{-0.2}$ DTC | $-3.1^{+0.3}_{-0.3}$ DTC | $-3.1^{+0.3}_{-0.3}$ DTC | $-3.2^{+0.2}_{-0.2}$ DTC | $-3.1^{+0.3}_{-0.3}$ DTC | $-3.1^{+0.3}_{-0.3}$ DTC | $-3.2^{+0.2}_{-0.2}$ DTC |
| CO (50 ppm) | NC | NC | NC | $-0.7^{+0.5}_{-0.6}$ UL | $-1.2^{+0.5}_{-0.5}$ UL | $-1.7^{+0.4}_{-0.3}$ UL | $-0.7^{+0.5}_{-0.5}$ UL | $-1.2^{+0.4}_{-0.4}$ UL | $-1.7^{+0.3}_{-0.4}$ UL | $-0.7^{+0.5}_{-0.5}$ UL | $-1.3^{+0.5}_{-0.4}$ UL | $-1.7^{+0.4}_{-0.4}$ UL | $-0.8^{+0.5}_{-0.5}$ UL | $-1.3^{+0.5}_{-0.4}$ UL | $-1.7^{+0.4}_{-0.4}$ UL |
| **SUMMARY:** | **Insufficient for characterizing Earth-thru time.** | | | **Marginal for characterizing Earth-thru time.** | | | **Sufficient for characterizing Earth-thru time.** | | | **Sufficient for characterizing Earth-thru time.** | | | **Sufficient for characterizing Earth-thru time.** | | |

**FIG. 4.** Summary of simulated retrievals showing the extent to which different long-wavelength cutoffs enable characterization of Earth-through-time biosignatures. Columns denote longwave cutoffs and assumed SNRs (R = 7 UV, R = 140 VIS, R = 70 NIR), whereas rows represent different atmospheric compositions assumed for Earth through time (true abundances in brackets). Each grid cell denotes the $\log_{10}$(mixing ratio) abundance constraint from a reflected light retrieval using the *rfast* spectral retrieval model (Robinson and Salvador, 2023). Green grid cells show detections with abundance constraints (*i.e.*, detections, denoted "DTC"), yellow grid cells represent upper limits ("UL") that are useful for contextualizing biosignatures or ruling out known false positive scenarios, whereas red grid cells represent either no constraints ("NC") or upper limits ("UL") that provide no useful contextual information for ruling out biosignature false positives or assessing habitability. $1\sigma$ uncertainties are reported for all gas detections or upper limits (where uncertainties in upper limits are uncertainties in the upper bound to the fitted analytic function). We find a long-wavelength cutoff of 1.7 $\mu$m is necessary for contextualizing Earth-like biosignatures throughout Earth's evolution.

| Long wavelength cutoff (μm): | 1.0 | | | 1.5 | | | 1.6 | | | 1.7 | | | 1.8 | | | 2.0 | | |
|---|---|---|---|---|---|---|---|---|---|---|---|---|---|---|---|---|---|---|
| SNR | 10 | 20 | 40 | 10 | 20 | 40 | 10 | 20 | 40 | 10 | 20 | 40 | 10 | 20 | 40 | 10 | 20 | 40 |
| **Proterozoic and Phanerozoic false positive scenarios** | | | | | | | | | | | | | | | | | | |
| **Photochemical oxygen false positives** | Oxygen detection completely ambiguous | | | Ambiguous as CO unconstrained but $H_2O$ upper limit | | | Sufficient to identify $O_2$ false positive via CO and low $H_2O$ | | | Sufficient to identify $O_2$ false positive via CO and low $H_2O$ | | | Sufficient to identify $O_2$ false positive via CO and low $H_2O$ | | | Sufficient to identify $O_2$ false positive via CO and low $H_2O$ | | |
| $CO_2$ (46%) | NC | NC | NC | $-0.6^{+0.3}_{-0.5}$ DTC | $-0.5^{+0.3}_{-0.3}$ DTC | $-0.4^{+0.2}_{-0.2}$ DTC | $-0.6^{+0.3}_{-0.5}$ DTC | $-0.6^{+0.3}_{-0.3}$ DTC | $-0.5^{+0.1}_{-0.2}$ DTC | $-0.6^{+0.3}_{-0.4}$ DTC | $-0.6^{+0.2}_{-0.3}$ DTC | $-0.5^{+0.1}_{-0.2}$ DTC | $-0.6^{+0.3}_{-0.4}$ DTC | $-0.6^{+0.2}_{-0.3}$ DTC | $-0.5^{+0.1}_{-0.2}$ DTC | $-0.6^{+0.3}_{-0.4}$ DTC | $-0.6^{+0.2}_{-0.3}$ DTC | $-0.5^{+0.1}_{-0.2}$ DTC |
| $H_2O$ (0.1 ppm) | $-4.6^{+0.7}_{-0.7}$ UL | $-4.9^{+0.5}_{-0.7}$ UL | $-5.3^{+0.4}_{-0.3}$ UL | $-5.6^{+0.5}_{-0.5}$ UL | $-5.8^{+0.4}_{-0.5}$ UL | $-5.8^{+0.3}_{-0.3}$ UL | $-5.8^{+0.6}_{-0.7}$ UL | $-5.9^{+0.4}_{-0.4}$ UL | $-5.9^{+0.2}_{-0.2}$ UL | $-5.8^{+0.6}_{-0.7}$ UL | $-5.8^{+0.4}_{-0.3}$ UL | $-5.9^{+0.2}_{-0.3}$ UL | $-5.7^{+0.6}_{-0.6}$ UL | $-5.9^{+0.4}_{-0.4}$ UL | $-6.0^{+0.2}_{-0.3}$ UL | $-5.9^{+0.5}_{-0.6}$ UL | $-6.1^{+0.4}_{-0.4}$ UL | $-6.3^{+0.3}_{-0.3}$ UL |
| CO (36%) | NC | NC | NC | $-0.4^{+0.3}_{-0.3}$ UL | $-0.4^{+0.3}_{-0.2}$ UL | $-0.4^{+0.3}_{-0.2}$ UL | $-0.3^{+0.2}_{-0.2}$ UL | $-0.3^{+0.2}_{-0.2}$ UL | $-0.7^{+0.2}_{-0.3}$ DTC | $-0.3^{+0.3}_{-0.2}$ UL | $-0.3^{+0.2}_{-0.1}$ UL | $-0.7^{+0.2}_{-0.3}$ DTC | $-0.3^{+0.3}_{-0.2}$ UL | $-0.3^{+0.2}_{-0.2}$ UL | $-0.7^{+0.2}_{-0.3}$ DTC | $-0.3^{+0.3}_{-0.2}$ UL | $-0.3^{+0.2}_{-0.2}$ UL | $-0.7^{+0.2}_{-0.3}$ DTC |
| **Waterworld oxygen false positives** | Sufficient to put crude upper limit on land fraction (LF) | | | Sufficient to put crude upper limit on land fraction (LF) | | | Sufficient to put crude upper limit on land fraction (LF) | | | Sufficient to put crude upper limit on land fraction (LF) | | | Sufficient to put crude upper limit on land fraction (LF) | | | Sufficient to put crude upper limit on land fraction (LF) | | |
| Land fraction, LF (0%) | LF< 28% | LF< 17% | LF< 11% | LF< 34% | LF< 27% | LF< 10% | LF< 35% | LF< 17% | LF< 10% | LF< 27% | LF< 17% | LF< 10% | LF< 25% | LF< 15% | LF< 9% | LF< 27% | LF< 15% | LF< 9% |
| Land fraction, LF (30%) | NC | $0.34^{+0.1}_{-0.1}$ DTC | $0.30^{+0.1}_{-0.1}$ DTC | NC | $0.31^{+0.2}_{-0.1}$ DTC | $0.28^{+0.1}_{-0.1}$ DTC | NC | $0.30^{+0.2}_{-0.1}$ DTC | $0.29^{+0.1}_{-0.1}$ DTC | NC | $0.31^{+0.1}_{-0.1}$ DTC | $0.29^{+0.1}_{-0.1}$ DTC | NC | $0.29^{+0.1}_{-0.1}$ DTC | $0.28^{+0.1}_{-0.1}$ DTC | NC | $0.28^{+0.1}_{-0.1}$ DTC | $0.27^{+0.1}_{-0.1}$ DTC |
| **Low non-condensable oxygen false positive** | Insufficient to identify low non-condensable background | | | Insufficient to identify low non-condensable background | | | Possible to identify false positive with sufficient SNR | | | Possible to identify false positive with sufficient SNR | | | Possible to identify false positive with sufficient SNR | | | Possible to identify false positive with sufficient SNR | | |
| $O_2$ (99.5%) | > -.46 DTC | >-.27 DTC | >-.10 DTC | >-.33 DTC | >-.12 DTC | >-.09 DTC | >-.12 DTC | >-.06 DTC | >-.05 DTC | >-.11 DTC | >-.05 DTC | >-.03 DTC | >-.16 DTC | >-.06 DTC | >-.04 DTC | >-.17 DTC | >-.06 DTC | >-.06 DTC |
| $N_2$ (0.01 ppm) | -0.63 UL | -0.73 UL | -1.22 UL | -0.74 UL | -0.94 UL | -1.26 UL | -0.85 UL | -1.11 UL | -1.09 UL | -0.82 UL | -1.08 UL | -1.32 UL | -0.64 UL | -1.05 UL | -1.15 UL | -0.72 UL | -1.14 UL | -0.94 UL |
| $P_{surf}$ (0.2 bar) | $4.4^{+0.3}_{-0.3}$ DTC | $4.3^{+0.2}_{-0.2}$ DTC | $4.3^{+0.1}_{-0.1}$ DTC | $4.6^{+0.4}_{-0.4}$ DTC | $4.5^{+0.3}_{-0.3}$ DTC | $4.4^{+0.2}_{-0.1}$ DTC | $4.6^{+0.4}_{-0.4}$ DTC | $4.5^{+0.3}_{-0.3}$ DTC | $4.4^{+0.2}_{-0.2}$ DTC | $4.6^{+0.4}_{-0.4}$ DTC | $4.5^{+0.3}_{-0.3}$ DTC | $4.4^{+0.1}_{-0.1}$ DTC | $4.6^{+0.4}_{-0.4}$ DTC | $4.5^{+0.3}_{-0.3}$ DTC | $4.4^{+0.1}_{-0.1}$ DTC | $4.6^{+0.4}_{-0.4}$ DTC | $4.5^{+0.3}_{-0.2}$ DTC | $4.4^{+0.1}_{-0.1}$ DTC |
| **Archean false positive scenarios** | | | | | | | | | | | | | | | | | | |
| **Reduced mantle methane false positive** | Completely insufficient to contextualize methane | | | Insufficient to identify CO-rich atmosphere | | | Marginal to identify CO-rich atmosphere | | | Marginal to identify CO-rich atmosphere | | | Marginal to identify CO-rich atmosphere | | | Marginal to identify CO-rich atmosphere | | |
| $CO_2$ (10%) | NC | NC | NC | $-0.8^{+0.5}_{-0.4}$ DTC | $-0.8^{+0.4}_{-0.3}$ DTC | $-0.9^{+0.3}_{-0.2}$ DTC | $-0.7^{+0.4}_{-0.5}$ DTC | $-0.8^{+0.4}_{-0.3}$ DTC | $-0.9^{+0.3}_{-0.2}$ DTC | $-0.7^{+0.4}_{-0.4}$ DTC | $-0.8^{+0.4}_{-0.3}$ DTC | $-0.9^{+0.2}_{-0.2}$ DTC | $-0.7^{+0.4}_{-0.4}$ DTC | $-0.8^{+0.4}_{-0.3}$ DTC | $-0.9^{+0.2}_{-0.2}$ DTC | $-0.7^{+0.4}_{-0.4}$ DTC | $-0.8^{+0.4}_{-0.3}$ DTC | $-0.9^{+0.2}_{-0.2}$ DTC |
| $CH_4$ (0.1%) | $-3.1^{+0.3}_{-0.2}$ DTC | $-3.2^{+0.3}_{-0.1}$ DTC | $-3.1^{+0.3}_{-0.1}$ DTC | $-3.0^{+0.4}_{-0.2}$ DTC | $-2.8^{+0.4}_{-0.3}$ DTC | $-2.9^{+0.3}_{-0.2}$ DTC | $-2.7^{+0.4}_{-0.3}$ DTC | $-2.8^{+0.4}_{-0.3}$ DTC | $-2.9^{+0.2}_{-0.2}$ DTC | $-2.7^{+0.4}_{-0.3}$ DTC | $-2.8^{+0.4}_{-0.3}$ DTC | $-2.9^{+0.2}_{-0.2}$ DTC | $-2.7^{+0.4}_{-0.3}$ DTC | $-2.8^{+0.4}_{-0.3}$ DTC | $-2.9^{+0.2}_{-0.2}$ DTC | $-2.7^{+0.4}_{-0.3}$ DTC | $-2.8^{+0.4}_{-0.3}$ DTC | $-2.9^{+0.2}_{-0.2}$ DTC |
| CO (1%) | NC | NC | NC | NC | NC | NC | $-0.6^{+0.5}_{-0.6}$ UL | $-1.0^{+0.4}_{-0.4}$ UL | $-1.5^{+0.3}_{-0.2}$ UL | $-0.6^{+0.5}_{-0.7}$ UL | $-1.1^{+0.4}_{-0.3}$ UL | $-1.5^{+0.3}_{-0.3}$ UL | $-0.6^{+0.4}_{-0.6}$ UL | $-1.1^{+0.4}_{-0.4}$ UL | $-1.5^{+0.3}_{-0.2}$ UL | $-0.6^{+0.5}_{-0.5}$ UL | $-1.0^{+0.4}_{-0.4}$ UL | $-1.5^{+0.3}_{-0.3}$ UL |
| **Hydrothermal methane false positive** | False positive not detectable-unproblematic | | | False positive not detectable-unproblematic | | | False positive not detectable-unproblematic | | | False positive not detectable-unproblematic | | | False positive not detectable-unproblematic | | | False positive not detectable-unproblematic | | |
| $CO_2$ (0.3%) | NC | NC | NC | $-1.7^{+0.6}_{-0.5}$ UL | $-1.9^{+0.3}_{-0.2}$ UL | $-2.2^{+0.2}_{-0.2}$ UL | $-1.7^{+0.6}_{-0.4}$ UL | $-1.9^{+0.3}_{-0.3}$ UL | $-2.3^{+0.3}_{-0.1}$ UL | $-1.7^{+0.6}_{-0.4}$ UL | $-2.1^{+0.4}_{-0.2}$ UL | $-2.6^{+0.2}_{-0.2}$ DTC | $-1.7^{+0.5}_{-0.4}$ UL | $-2.1^{+0.4}_{-0.2}$ UL | $-2.6^{+0.2}_{-0.2}$ DTC | $-1.6^{+0.5}_{-0.4}$ UL | $-2.1^{+0.3}_{-0.2}$ UL | $-2.6^{+0.2}_{-0.2}$ DTC |
| $CH_4$ (1.6 ppm) | $-4.1^{+0.5}_{-0.4}$ UL | $-4.2^{+0.5}_{-1.4}$ UL | $-4.0^{+0.5}_{-1.2}$ UL | $-4.0^{+0.6}_{-0.6}$ UL | $-4.2^{+0.5}_{-0.5}$ UL | $-4.6^{+0.3}_{-0.3}$ UL | $-3.6^{+0.5}_{-0.5}$ UL | $-4.2^{+0.4}_{-0.4}$ UL | $-4.6^{+0.3}_{-0.3}$ UL | $-4.4^{+0.5}_{-0.5}$ UL | $-4.8^{+0.4}_{-0.4}$ UL | $-5.0^{+0.3}_{-0.2}$ UL | $-4.4^{+0.5}_{-0.4}$ UL | $-4.8^{+0.4}_{-0.3}$ UL | $-5.1^{+0.3}_{-0.2}$ UL | $-4.4^{+0.6}_{-0.5}$ UL | $-4.8^{+0.4}_{-0.4}$ UL | $-5.0^{+0.3}_{-0.2}$ UL |
| CO (0.3 ppm) | NC | NC | NC | NC | NC | NC | $-1.1^{+0.5}_{-0.5}$ UL | $-1.5^{+0.3}_{-0.3}$ UL | $-1.9^{+0.3}_{-0.2}$ UL | $-1.1^{+0.6}_{-0.3}$ UL | $-1.7^{+0.4}_{-0.3}$ UL | $-2.2^{+0.4}_{-0.4}$ UL | $-1.1^{+0.6}_{-0.5}$ UL | $-1.6^{+0.4}_{-0.4}$ UL | $-2.2^{+0.4}_{-0.3}$ UL | $-1.2^{+0.6}_{-0.5}$ UL | $-1.8^{+0.5}_{-0.4}$ UL | $-2.2^{+0.4}_{-0.4}$ UL |
| **SUMMARY:** | **Known false positive cannot be identified** | | | **Known false positives cannot be identified** | | | **Sufficient to identify most false positives** | | | **Sufficient to identify most false positives** | | | **Sufficient to identify most false positives** | | | **Sufficient to identify most false positives** | | |

**FIG. 5.** Summary of simulated retrievals showing the extent to which different long-wavelength cutoffs enable identification of Earth-through-time biosignature false positives. Columns denote longwave cutoffs and simulated SNRs (R = 7 UV, R = 140 VIS, R = 70 NIR), whereas rows represent different biosignature false positive scenarios and their assumed atmospheric (or surface) compositions. True abundances are shown in brackets—see Table 3 for details. Each grid cell denotes the $\log_{10}$(mixing ratio) abundance constraint from a reflected light retrieval using the *rfast* spectral retrieval model (Robinson and Salvador, 2023). Green grid cells show detections with abundance constraints ("DTC"), yellow grid cells represent upper limits ("UL") that are useful for contextualizing biosignatures or ruling out known false positive scenarios, whereas red grid cells represent either no constraints ("NC") or upper limits ("UL") that provide no useful contextual information for ruling out biosignature false positives or assessing habitability. $1\sigma$ uncertainties are reported for gas detections or upper limits (where uncertainties in upper limits are uncertainties in the upper bound to the fitted analytic function). Uncertainties are not cited for a handful of unambiguous detections and upper limits that are not well-fitted by analytic functions; we instead quote 95% credible limits. We find a long-wavelength cutoff of ≥1.6 $\mu$m and high SNR (20–40) are necessary for identifying most biosignature false positive scenarios.

oxygen false-positive scenario from an inhabited Earth-analog with a $\sim 10^{-3}$ water mixing ratio.

The waterworld surface retrievals follow the approach by Ulses et al. (2025) except that the "true" planetary surface is 100% ocean. Waterworld oxygen false positives can never be unambiguously identified by reflectance spectra alone since it is impossible to rule out small islands in a shallow ocean in a disk-averaged spectrum. However, it is possible to put a robust upper limit on the land fraction of the visible hemisphere of the planet, which might be suggestive of a waterworld false positive if oxygen is high but no other corroborating biosignatures ($CH_4$, $N_2O$, etc.) are observed. To more clearly illustrate the ability of reflected light observations to rule out waterworlds, we also include simulated retrievals of Earth analogs (i.e., 30% land) to explore the conditions required for land detection. While the NIR cutoff has virtually no impact on land fraction constraints, high SNR (20–40) would detect Earth-like land or provide land fraction upper limits much lower than that of modern Earth; the former would preclude a waterworld, whereas the latter could indicate a waterworld false positive as a possible explanation for atmospheric oxygen.

For low noncondensable oxygen false positives, the most important requirement is sufficient SNR to constrain background pressure. At a minimum, SNR 20 is needed to obtain an order of magnitude surface pressure constraint, whereas SNR 40 would enable a much more precise total pressure and $pO_2$ constraint, thereby identifying the lack of noncondensable gases as a plausible explanation for any observed $pO_2$.

For methane false-positive scenarios, we find that hydrothermal methane false positives are unproblematic because the abiotic methane is not plausibly detectable for any SNR we considered (and thus cannot be mistaken for biogenic methane). For the reduced mantle methane degassing scenario, wavelength coverage to at least 1.6 μm is needed to both constrain $CO_2$ and provide an upper limit on CO. With that said, the reduced methane false-positive scenario is particularly challenging to identify with bulk abundances alone since, even at SNR 40, CO is not detected, so the $CH_4$:CO ratio—which is potentially diagnostic of such false positives (Thompson et al., 2022)—is poorly constrained. Further work is needed to investigate the long-term stability of atmospheres rich in CO, $CO_2$, and $CH_4$ because carbon cycling on planets that possess sufficiently reducing mantles to degas $CH_4$ may operate differently than on Earth. Overall, the long-wavelength requirements necessary to identify false-positive scenarios are comparable to those for identifying genuine biosignature scenarios, with the added requirement that high SNR is needed to definitively identify low noncondensable false positives. Such high SNRs are likely to only be sought on already-promising targets.

### 3.2. Short-wavelength cutoff

Next, we examine the short-wavelength cutoff needed to confidently identify oxygen and methane biosignatures and disfavor known nonbiological false positives. Figure 6 summarizes results from Proterozoic Earth retrievals (0.1% $O_2$ PAL, low $CH_4$) whereby the short-wavelength cutoff has been systematically varied from 0.2 to 0.415 μm (SNR = 10, defined at 0.366 continuum as described in the Supplementary Data). The resulting constraints on ozone are plotted alongside the assumed Proterozoic ozone abundance ($1 \times 10^{-8}$). Broadly speaking, we find a short-wavelength cutoff of no longer than 0.26 μm is needed to provide a confident ozone detection, thereby identifying an inhabited Proterozoic Earth-analog that might otherwise be a false negative. Note that biogenic $CH_4$ is undetectable for any plausible SNR and long-wavelength cutoff in this low $CH_4$ scenario, as shown in Figure 4, emphasizing the importance of detecting $O_3$ as a biosignature for such planets.

Figure 7 shows biosignature interpretation results of spectral retrieval analyses for the short-wavelength coronagraph cutoff. Specifically, we show $O_3$ gas constraints for Phanerozoic and Proterozoic scenarios for a range of short-wavelength cutoffs and assumed SNRs (Archean cases are neglected since they do not have atmospheric ozone). Unlike the longwave cutoff calculations described above, where we are interested in constraining the full planetary context, these retrievals are limited to UV spectra (X–0.45 μm, where X = 0.2, 0.24, 0.25, 0.264, 0.288, or 0.346 μm) to be most conservative with respect to ozone detectability, for example, ignoring information from the 0.4–0.65 μm Chappuis band. Consistent with the calculations in Figure 6, we find that UV coverage to at least 0.26 μm is necessary to characterize ozone abundances throughout Earth's (oxygenated) history and that shorter wavelength coverage provides only incrementally improved ozone abundances. While a cutoff of 0.288 μm may provide an ozone detection given high SNR observations (SNR ≥40), achieving such a high SNR in the UV would likely require prohibitive integration times given low stellar fluxes and the lack of planetary reflectivity in $O_3$ absorption bands, especially considering the much lower SNR needed when shorter UV measurements are available. Similarly, a cutoff of 0.288 μm might enable detection of modern Earth-like $O_2$/$O_3$ levels, but atmospheric $O_2$ is easily detectable at visible wavelengths in such atmospheres, so a short-wavelength cutoff of 0.288 μm (or longer) would not substantially improve biosignature detection capabilities over visible wavelength characterization.

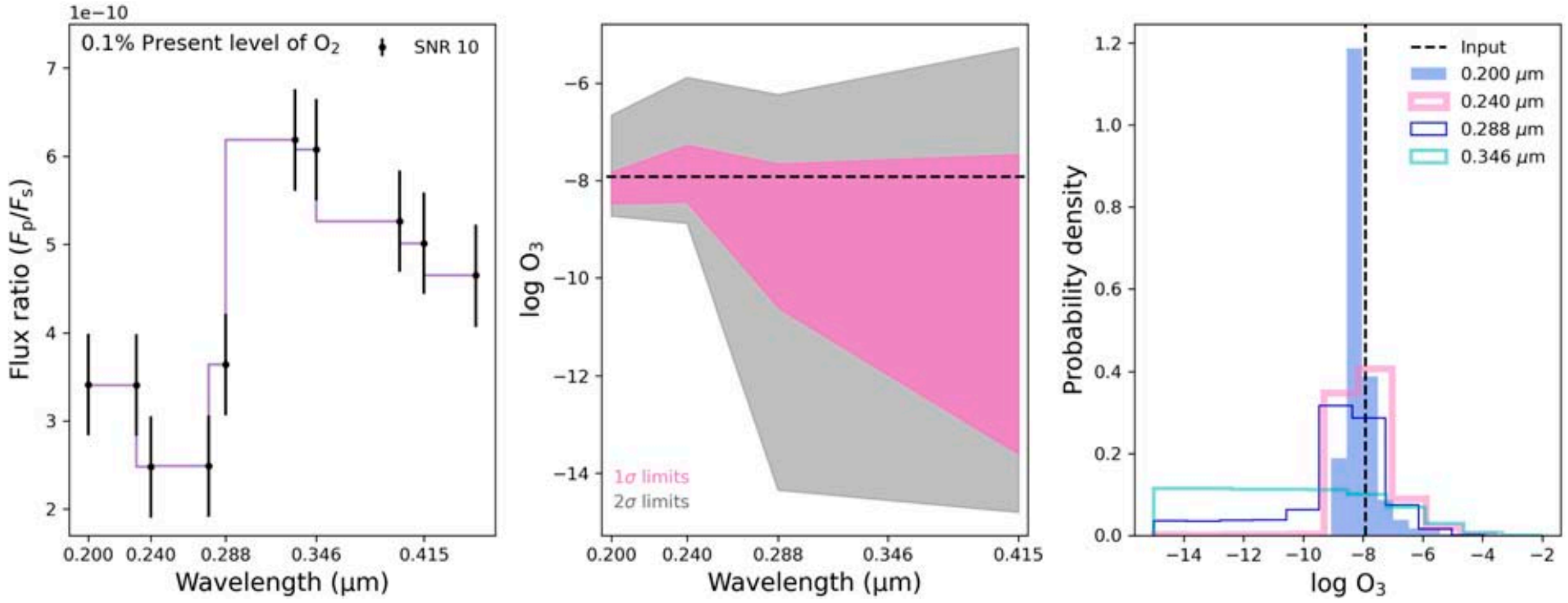

**FIG. 6.** Short-wavelength cutoff required for identifying Proterozoic Earth biosignatures via ozone detection. Left subplot denotes full UV spectrum and assumed uncertainties. Middle subplot shows the $1\sigma$ (purple) and $2\sigma$ (gray) uncertainties on ozone abundances for the full range of assumed short-wavelength cutoffs, along with the true column-averaged value (dashed vertical line). The right subplot shows the ozone mixing ratio posterior for selected shortwave cutoffs. Robust Proterozoic $O_3$ detection requires a shortwave cutoff of around 0.26 μm.

| Short wavelength cutoff (μm): | 0.200 | | | 0.240 | | | 0.250 | | | 0.264 | | | 0.288 | | | 0.346 | | |
|---|---|---|---|---|---|---|---|---|---|---|---|---|---|---|---|---|---|---|
| SNR | 10 | 20 | 40 | 10 | 20 | 40 | 10 | 20 | 40 | 10 | 20 | 40 | 10 | 20 | 40 | 10 | 20 | 40 |
| **Phanerozoic Earth** (0-0.54 Ga) | Sufficient to identify $O_3$ biosig. | | | Sufficient to identify $O_3$ biosig. | | | Sufficient to identify $O_3$ biosig. | | | Sufficient to identify $O_3$ biosig. | | | Sufficient to identify $O_3$ biosig. | | | Insufficient to identify $O_3$ biosig. | | |
| $O_3$ (100% PAL $O_2$) | $-5.9^{+1.1}_{-0.8}$ DTC | $-6.1^{+0.9}_{-0.6}$ DTC | $-6.1^{+0.6}_{-0.4}$ DTC | $-6.4^{+1.1}_{-0.6}$ DTC | $-6.1^{+0.9}_{-0.5}$ DTC | $-5.8^{+0.8}_{-0.5}$ DTC | $-6.5^{+1.1}_{-0.6}$ DTC | $-6.2^{+1.0}_{-0.5}$ DTC | $-6.0^{+0.8}_{-0.5}$ DTC | $-6.4^{+1.2}_{-0.5}$ DTC | $-6.2^{+0.9}_{-0.5}$ DTC | $-5.9^{+0.8}_{-0.5}$ DTC | $-6.6^{+1.2}_{-0.6}$ DTC | $-6.2^{+1.0}_{-0.5}$ DTC | $-6.0^{+0.8}_{-0.5}$ DTC | -6.2 UL | -6.6 UL | -6.9 UL |
| **Proterozoic Earth high $O_3$** (0.54-2.5 Ga) | Sufficient to identify $O_3$ biosig. | | | Sufficient to identify $O_3$ biosig. | | | Sufficient to identify $O_3$ biosig. | | | Sufficient to identify $O_3$ biosig. | | | Insufficient to identify $O_3$ biosig. | | | Insufficient to identify $O_3$ biosig. | | |
| $O_3$ (1% PAL $O_2$) | $-6.5^{+0.9}_{-0.6}$ DTC | $-6.5^{+0.7}_{-0.4}$ DTC | $-6.3^{+0.6}_{-0.4}$ DTC | $-6.6^{+1.0}_{-0.6}$ DTC | $-6.5^{+0.9}_{-0.4}$ DTC | $-6.2^{+0.7}_{-0.5}$ DTC | $-6.7^{+1.0}_{-0.6}$ DTC | $-6.6^{+0.8}_{-0.4}$ DTC | $-6.4^{+0.8}_{-0.5}$ DTC | $-6.6^{+1.0}_{-0.6}$ DTC | $-6.5^{+0.8}_{-0.4}$ DTC | $-6.1^{+0.7}_{-0.5}$ DTC | -5.1 UL | -5.2 UL | $-6.4^{+0.8}_{-0.5}$ DTC | -6.1 UL | -6.5 UL | -6.5 UL |
| **Proterozoic Earth low $O_3$** (0.54-2.5 Ga) | Sufficient to identify $O_3$ biosig. | | | Sufficient to identify $O_3$ biosig. | | | Sufficient to identify $O_3$ biosig. | | | Sufficient to identify $O_3$ biosig. | | | Insufficient to identify $O_3$ biosig. | | | Insufficient to identify $O_3$ biosig. | | |
| $O_3$ (0.1% PAL $O_2$) | $-8.2^{+0.4}_{-0.3}$ DTC | $-8.1^{+0.3}_{-0.2}$ DTC | $-7.9^{+0.4}_{-0.2}$ DTC | $-8.1^{+0.8}_{-0.4}$ DTC | $-7.9^{+0.8}_{-0.3}$ DTC | $-7.8^{+0.6}_{-0.3}$ DTC | $-7.9^{+0.9}_{-0.4}$ DTC | $-7.7^{+0.9}_{-0.3}$ DTC | $-7.4^{+0.6}_{-0.3}$ DTC | $-7.9^{+0.9}_{-0.4}$ DTC | $-7.8^{+0.8}_{-0.3}$ DTC | $-7.5^{+0.7}_{-0.3}$ DTC | -6.6 UL | -6.4 UL | $-7.8^{+1.0}_{-0.4}$ DTC | -6.0 UL | -6.4 UL | -6.5 UL |
| **SUMMARY:** | Sufficient for characterizing Earth-thru time. | | | Sufficient for characterizing Earth-thru time. | | | Sufficient for characterizing Earth-thru time. | | | Sufficient for characterizing Earth-thru time. | | | Insufficient for characterizing Earth-thru time. | | | Insufficient for characterizing Earth-thru time. | | |

**FIG. 7.** Summary of simulated retrievals showing the extent to which different short-wavelength cutoffs enable characterization of Earth-through-time biosignatures. Columns denote shortwave cutoffs and input spectra SNRs (R = 7 UV), whereas rows represent different atmospheric compositions assumed for Earth through time. Each grid cell denotes the $\log_{10}$(mixing ratio) $O_3$ abundance constraint from a reflected light retrieval using the *rfast* spectral retrieval model (Robinson and Salvador, 2023). Green grid cells show detections with abundance constraints ("DTC"), whereas red grid cells represent 95% upper limits ("UL") that provide no useful contextual information for identifying an $O_3$ biosignature. We find a short-wavelength cutoff of 0.26 $\mu$m is necessary for contextualizing Earth-like biosignatures throughout Earth's evolution.

### *3.3. Summary of requirements for biosignature and false-positive identification*

Figure 8 summarizes the longwave cutoff results for all the Earth-like biosignature and false-positive scenarios considered in this study. For each NIR wavelength cutoff, we have tallied the number of scenarios that can be positively identified—each of the cases and SNR combinations in Figures 4, 5, and 7 are considered unique scenarios (see the Supplementary Data for details). Positive identification implies that either (1) for Earth-through-time scenarios, biosignature gases can be detected and all confounding false-positive cases can be tentatively excluded based on other atmospheric constraints, or (2) for false-positive cases, the false-positive scenario can be identified by atmospheric constraints, and there is no potential ambiguity with genuine Earth-through-time biosignature scenarios. The calculation was repeated assuming with and without UV coverage to 0.26 μm, the main difference being the inability to identify Proterozoic-like biospheres, as well as waterworld false positives, as described by Ulses et al. (2025). While this fractional approach is crude and scenario-dependent, it highlights the primary importance of long-wavelength capabilities to at least 1.7 μm, in addition to UV capabilities to constrain $O_3$.

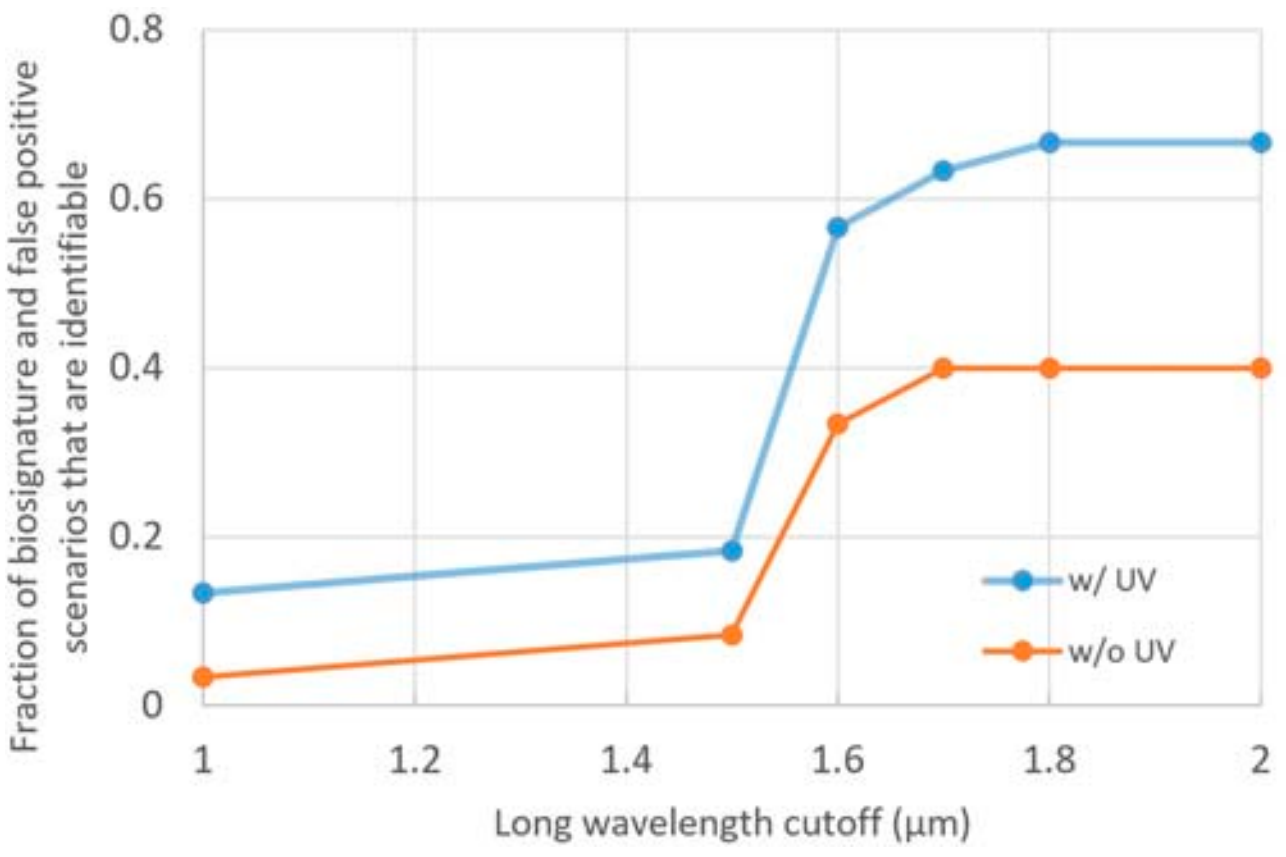

**FIG. 8.** Fraction of biosignature and biosignature false positive scenarios in this study that can be characterized as a function of long-wavelength cutoff. Results are shown both with (blue line) and without (orange line) UV coverage. Scenarios include all the Earth-through-time biosignature scenarios in Figures 4 and 7, as well as false positive scenarios in Figure 5, and each assumed SNR value counts as a separate case (30 scenarios total). NIR coverage to at least 1.7 μm and UV coverage to no greater than 0.26 μm will enable identification of the vast majority of biosignature and false positive scenarios. See the Supplementary Data for additional information on how this figure was generated.

## 4. Implications for Telescope Design

### 4.1. Wavelength range and SNR requirements for life detection

The Astrophysics Decadal Survey calls for a UV/VIS/NIR telescope to search for signs of life on ~25 potentially Earth-like planets. To meet this goal, the HWO telescope must be able to both identify Earth-like biosignatures and rule out plausible biosignature false positives for the bulk of these potential exo-Earths. Given the simulated reflected light retrievals investigated in this study, we conclude HWO coronagraphs must be able to obtain 0.26–1.7 μm spectra for most targets to conduct a statistically meaningful search for biosignatures. Moreover, it must be possible to obtain high SNR (20–40) for any of the ~25 targets (assuming R = 7 UV; R = 140 VIS; R = 70 NIR) to follow up on potential biosignatures and more fully characterize the planetary context.

We recommend a wavelength cutoff of 1.7 μm rather than 1.6 μm because, while these two cutoffs provide comparable information at high SNR, a 1.7-μm cutoff gives valuable additional information on planetary context at lower SNR values (SNR = 10). Crucially, obtaining high SNR (≥20) all the way to the edge of the NIR spectrum may be challenging given the increasing thermal background (see below), so a cutoff of 1.7 μm is a less risky choice. Extending spectral coverage to longer wavelengths (>1.7 $\mu$m) to measure additional species not considered here could provide additional opportunities to identify corroborating biosignatures and contextual clues on a handful of planets observable at those wavelengths (*i.e.*, brighter targets), but these longer wavelength capabilities need not drive observatory thermal design given the challenges associated with a cold (<0°C) telescope.

We recommend a shortwave cutoff of 0.26 μm rather than 0.28–0.29 μm because the latter may preclude detection of life on Proterozoic-like Earth analogs, which constitute about half of Earth's inhabited history. While UV starlight suppression to $\sim 10^{-10}$ is undoubtedly a major technical challenge (Balasubramanian et al., 2011; Pueyo et al., 2019), we note that the assumed UV resolution (R ~ 7) is low, requiring only a handful of photometric datapoints, which will help ensure feasible integration times even if UV throughput is low.

The spectral capabilities outlined above would provide sufficient information to step through the standards of evidence required for life detection (Meadows et al., 2022). Achieving SNR ≥20 across the full wavelength range may take considerable observing time with HWO; a graduated approach could be taken, where high SNR observations are only sought on already-promising candidates that show signs of habitability and possible biosignatures at lower SNRs or over more restricted wavelengths (Young et al., 2024a).

### 4.2. Implications for HWO aperture and number of accessible targets

While detailed yield studies are beyond the scope of this study, the combination of wavelength and SNR requirements described above suggests that a large aperture (*i.e.*, >8 m) will be necessary to provide the IWA and feasible integration times necessary for identifying and interpreting biosignatures for a sample of ~25 habitable zone rocky planets (Morgan et al., 2024). A 6-m telescope would likely only be able to characterize ~5 exo-Earths with NIR spectra extending to 1.7 μm (Morgan et al., 2023). Improved coronagraph throughput and coronagraph IWA are also important to improve planet characterization prospects. HWO yield modeling suggests that a combination of large aperture (>8 m), aggressive IWA (<30 mas), and high contrast ($<10^{-10}$) would be necessary to characterize 15–20 exo-Earth candidates out to 1.65 $\mu$m (Morgan et al., 2024).

## 5. Discussion

### 5.1. Comparisons to previous work

Here, we compare the key findings of this study to other independent retrieval studies that have investigated aspects of Earth-analog detectability and its wavelength dependence. The purpose of these comparisons is to emphasize that the wavelength cutoff recommendations described above are not anomalous but are instead broadly consistent with this body of prior work. This consistency across independent studies strengthens our recommendations for telescope design and is a step toward building a consensus view on Earth-analog characterization requirements.

The retrievals presented in this study corroborate and expand upon the results of Damiano and Hu (2022), who conducted reflected light retrievals with (0.4–1.8 μm) and without (0.4–1.0 μm) the NIR portion of the spectrum and found that NIR capabilities are essential for constraining the bulk atmosphere of Earth-through-time analogs. Our results also complement simulated retrievals of thermal emission observations (4–18.5 μm) to characterize Earth through time (Alei et al., 2022) with the proposed Large Interferometer for Exoplanets mission concept (Quanz et al., 2022).

The results of this study are broadly consistent with those of Tokadjian et al. (2024), where a different radiative transfer and retrieval code, EXOREL, was used to investigate biosignature detectability on Archean and Proterozoic Earth. This study similarly conducted reflected light retrievals assuming SNR = 20 across the UV to NIR (R = 7 UV, R = 140 VIS, and R = 70 NIR). It was found that changing the long-wavelength cutoff from 1.6 to 1.8 μm does not dramatically alter biogenic methane constraints, in agreement with the findings of our study (cf. our Fig. 4). Tokadjian et al. (2024) also found that CO detections would be challenging for SNR = 20, although upper limits may be inferred for low bulk abundances, as in our Figures 4 and 5. Finally, Tokadjian et al. (2024) investigated the detectability of Proterozoic $N_2O$, a potential biosignature gas (Schwieterman et al., 2022) not considered in this study, and found a long-wavelength cutoff of at least 1.4 μm is needed to detect abundant ($>10^{-3}$) $N_2O$.

The results of this study are also consistent with those of Latouf et al. (2025), who conducted retrievals over a large grid of $CH_4$, $O_2$, $O_3$, and $H_2O$ abundances broadly representative of Earth through time. This study was focused on determining the coronagraph bandpasses and SNRs required to distinguish $H_2O$ and $CH_4$, again using an entirely independent radiative transfer and retrieval scheme based on the Planetary Spectrum Generator (Villanueva et al., 2018) and *MultiNest* (Feroz et al., 2009). Latouf et al. (2025) found that Archean $CH_4$ abundances ought to be detectable regardless of long-wavelength cutoff and SNR (in agreement with Figs. 3 and 4), whereas modern Earth-like $CH_4$ is extremely challenging to detect (cf. Fig. 4). The same methodology was used to show that CO abundances up to 1% are not constrainable at SNR ≤15 given a 20% coronagraph bandpass centered anywhere between 0.8 and 2.0 μm, and that only weak CO constraints are possible for 1% CO and SNR >15 (Hagee et al., 2026); given the CO upper limits found in this study (Figs. 4 and 5), these results suggest that background gas constraints will likely require spectral observations at multiple bandpasses to determine the full spectrum. Hagee et al. (2026) also found that $CO_2$ abundances less than a Proterozoic level (*i.e.*, $1 \times 10^{-2}$ mixing ratio) are not detectable or constrainable at SNRs <20 and that, at abundances equal to or higher than a Proterozoic level, moderate SNRs ($\sim 15$ or lower) are sufficient to detect and constrain $CO_2$ at shorter wavelengths, as short as 1.5 μm (cf. Fig. 4), although the presence of $H_2O$ and $CH_4$ in this regime can also impact $CO_2$ detectability.

The results of this study are compatible with those of Gilbert-Janizek et al. (2024), where the retrieval code *smarter* was used to explore biosignature detectability and interpretation for a modern Earth twin with spectra generated using realistic noise for an 8-m, LUVOIR-B-like telescope. Their study is most comparable to our Phanerozoic Earth case. In agreement with the results of this work, they found that SNR $\sim$ 20 (SNR $\sim$ 17 in the visible, SNR $\sim$ 15 in the near-IR) and a longwave cutoff of 2.0 microns is sufficient to contextualize the $O_2/O_3$ biosignature, along with upper limits for the abundance of $CH_4$ and CO. While Gilbert-Janizek et al. (2024) found that $CO_2$ can be weakly constrained, this difference between the studies is likely caused by their slightly higher resolution of R = 100 in the near-IR. In this work, a $CO_2$ constraint is recovered when the SNR is increased to 40, supporting the conclusion that additional spectral information in this wavelength range may transform an upper limit into a weak detection.

We can also compare the results from the short-wavelength cutoff retrievals to prior work such as that by Damiano et al. (2023), where retrievals covering the UV to visible wavelength range were performed to evaluate the constraints able to be placed on $O_2$ and $O_3$ for 1% and 0.1% PAL of $O_2$ (Proterozoic Earth-like cases). Although our analog simulations focus exclusively on the UV, our findings are largely in agreement with those of Damiano et al. (2023), who identified a 0.250-micron cutoff as sufficient for robust $O_3$ detections. Our systematic evaluation of the short-wavelength cutoff extends this result further by demonstrating that the cutoff should not exceed 0.288 microns, as indicated by the extended statistical tail in the $O_3$ marginal posterior at this limit. These findings apply to both Proterozoic Earth cases and suggest that visible observations may not be strictly necessary for improving the $O_3$ constraints and that the UV alone would be sufficient to achieve strong detections. In contrast, Latouf et al. (2024) focused on $O_2/O_3$ detection in the visible and highlighted that constraining $O_3$ in the visible might require larger bandpass widths due to the broad extent of the $O_3$ Chappuis band. Our results show that even with nominal 20% bandpasses and moderate SNRs, $O_3$ constraints are more easily attainable in the UV. This further suggests that accessing the UV is critical, even for planetary atmospheres with high enough $O_3$ abundances to exhibit spectral features at longer wavelengths. It should also be noted that possible degeneracies between the $O_3$ Chappuis band and land surface spectra have not been fully incorporated in visible spectral retrievals, so UV observations may be necessary to break the $O_3$-land degeneracy (Ulses et al., 2025).

### 5.2. Caveats, limitations, and opportunities for future work

While these calculations provide useful heuristics for planning future observatories, several simplifying assumptions were made in the retrieval analysis. In particular, our results are limited by current knowledge of molecular opacities, planetary biogeochemistry, and photochemistry. The calculations presented in this paper conservatively assume $N_2$-like CIA for CO because CO–CO CIA data are not available. Future experiments or theoretical work to accurately compute CO pressure broadening could be incorporated into NIR-cutoff retrievals to determine whether CO may be more easily constrained than our calculations imply. This could also improve $CO_2$ constraints for high CO atmospheres given the overlap in $CO_2$ and CO absorption features around 1.5–1.6 μm. Additionally, the radiative transfer and retrieval model adopted here makes many simplifying assumptions including gray clouds, isothermal atmospheric structure, and well-mixed abundance profiles (for both true and retrieved atmospheres) that could either under- or overestimate detectability. For example, real water vapor profiles are nonuniform, which could result in underestimated water abundances for high cloud coverage. However, these uncertainties do not undermine our salient argument that broad NUV-VIS-NIR coverage is essential to life detection with HWO.

Our scenarios and corresponding gas abundances for Earth through time (and their false positives) are necessarily illustrative and not exhaustive. Here, we explore more specific limitations to our results as they apply to reducing and oxidizing atmospheres, in addition to practical observing limitations.

#### 5.2.1. Limits to identifying methane biosignatures and their false positives.

Arguably, the biggest obstacle to identifying Archean-like biosignatures is anticipating the precise relationship between $CH_4$, CO, and $CO_2$ for the full diversity of inhabited and lifeless worlds and then constraining these gas abundances with sufficient precision to disentangle these scenarios. While ≥1.7 $\mu$m NIR coverage enables crude upper limits on CO abundances, these upper limits may not always constrain the $CO/CH_4$ ratio well enough to rule out all false positives. Indeed, the ability to rule out the reduced mantle degassing scenario considered in this study is marginal given that the assumed CO abundance (1%) is compatible with the many inferred upper limits for biosignature cases (Figs. 3 and 4). With that said, the relationship between $CO_2$, CO, and $CH_4$ abundances for lifeless terrestrial planets has yet to be explored self-consistently for the full diversity of planetary compositions (e.g., Watanabe and Ozaki, 2024). The extremely reducing mantle conditions required to sustain high $CH_4$ fluxes may be incompatible with high $CO_2$ abundances sustained by a geologic carbon cycle. Similarly, the biogeochemical cycling of CO, $CH_4$, and $CO_2$ on inhabited worlds and the extent to which CO represents an uneaten "free lunch"—an "antibiosignature"—is also a topic of ongoing work (Schwieterman et al., 2019; Sauterey et al., 2020; Thompson et al., 2022; Akahori et al., 2024; Eager-Nash et al., 2024). Additionally, the flux–abundance relationship for Earth-like biosignatures will depend on the stellar spectral energy distribution (Segura et al., 2005; Arney, 2019), so biosphere detectability thresholds may differ across diverse HWO target stars. There is an opportunity to more thoroughly investigate the long-term geochemical plausibility of proposed methane degassing and hydrothermal false positives, as well as the full spectrum of $CH_4$-rich atmospheres compatible with surface biospheres.

Our study also ignores the possibility of volatile-rich, Titan-analog false positives for $CH_4$ biosignatures, as well as more exotic false positives such as late veneers producing compositionally distinct outer shells, thereby allowing for greater redox diversity in interior volatile fluxes (Kite, 2022; Thompson et al., 2022). The self-consistency of these scenarios has yet to be thoroughly explored, so their inclusion in this retrieval analysis would be premature. However, as more biosignature false-positive scenarios are developed and vetted, the necessary reflected light capabilities to observationally identify such scenarios ought to be systematically investigated.

#### 5.2.2. Limits to identifying oxygen biosignatures and their false positives.

While oxygen biosignatures and their false positives have been more thoroughly explored than those for methane (Meadows et al., 2018), future work ought to explore the long-term stability of low noncondensable false positives, as well as the detectability of runaway greenhouse $O_2$ false positives in reflected light. In general, $O_2$ false positives, whereby $O_2$ results from cumulative H escape (evolutionary false positives), have yet to be fully explored for planets around F/G/K stars (Krissansen-Totton et al., 2021), which will constitute the majority of the HWO target list (Mamajek and Stapelfeldt, 2024).

#### 5.2.3. The impact of fragmented spectra and bandpass offsets.

All the retrievals in this study assume continuous spectral coverage from the shortwave cutoff to the longwave cutoff, with fixed noise across the entire spectrum. In practice, noise profiles will be wavelength dependent, and planetary spectra obtained by HWO observations will need to be stitched together from multiple coronagraph bandpasses (Latouf et al., 2023; Young et al., 2024a; Latouf et al., 2025), whereby noise profiles will also vary with spectral distance from the bandpass center.

The challenges associated with stitching together spectral observations from multiple instruments, instrument modes, and detectors have already been encountered with JWST data; spectral offsets may need to be introduced in retrievals to accommodate fragmented spectra (Rustamkulov et al., 2023; Alderson et al., 2024). Moreover, while unambiguous spectral features are typically unaffected by offset assumptions, the detection significance of more subtle and/or broad molecular features may be strongly impacted. In the case of the transiting sub-Neptune K2-18b, large $CH_4$ features are robust to offset assumptions, but the detection significance of trace dimethylsulfide depends on detector offsets (Madhusudhan et al., 2023; Schmidt et al., 2025). For HWO, we similarly anticipate strong absorption features (*e.g.*, from $O_2$, $CH_4$, $CO_2$, $O_3$) to be robust to bandpass offsets unless features straddle multiple bandpasses, although the loss of information from the continuum may affect total pressure, cloud, and surface composition constraints. Naturally, simulated retrievals using more realistic, sequential coronagraph observations are a logical next step.

#### 5.2.4. The impact of spatial/temporal variability and viewing geometry.

All calculations in this study assume globally averaged 1D atmospheres observed at quadrature. This is both a limitation and an opportunity. Planetary variability over orbital phase may confound simple retrieval interpretations, especially if long integration times and sequential coronagraphic bandpasses necessitate observations that integrate over changing phases and planetary conditions. Conversely, 3D surface and cloud variability as well as phase-dependent information could be leveraged by time-resolved observations to improve constraints on planetary context (Robinson et al., 2010; Lustig-Yaeger et al., 2018; Cowan and Fujii, 2024). Future simulated observations ought to consider retrievals that integrate over such variability. Seasonal variations in atmospheric gases, particularly $O_3$, may also confound simple assessments of whether a planet is Earth-like but may ultimately be leveraged as corroborating evidence for life (Olson et al., 2018).

#### 5.2.5. Realistic thermal noise and the longwave cutoff.

It is also worth highlighting that simulating realistic noise is especially important for the long-wavelength cutoff, where thermal noise from the instruments could affect the ability to constrain CO around 1.6 μm and longer wavelengths. Indeed, increasing telescope temperature from 270 to 300 K has a dramatic influence on telescope noise beyond $\sim$1.65 μm (Hagee et al., 2026). In many respects, this work is optimistic regarding possible information gleaned from the edges of NIR spectra since we assumed a fixed SNR across the entire spectrum, whereas in practice noise will increase toward the edge of the NIR spectrum. Future work ought to investigate whether the high SNRs (20–40) required to constrain $CO/CO_2$ around 1.5–1.6 μm can be achieved for the majority of HWO targets with room temperature instruments, or whether passive cooling to 0°C will be necessary to enable robust biosignature interpretation. In the absence of such studies, a 1.7-μm cutoff is preferable to a 1.6-μm cutoff because it provides a buffer in the event that instrument performance deteriorates at the edge of the longest wavelength bandpass.

## 6. Conclusions

To conduct a meaningful search for life with the potential for robust conclusions with regard to planetary biosignatures and their interpretation, the HWO coronagraph(s) must have a wavelength range that extends from 0.26 to 1.7 μm for as many targets as possible. The shortwave limit is driven by $O_3$ detectability in weakly oxygenated atmospheres, and the longwave limit is driven by the need to constrain C-bearing molecules to disfavor biosignature false positives. Beyond 1.7 μm, there are diminishing returns for the purposes of characterizing oxygen and methane biosignatures, and any marginal benefits are likely confounded by increasingly noisy observations and IWA constraints. The ability to achieve a high SNR ( ∼ 20–40) as needed is desirable for ruling out many false positives, including photochemical oxygen false positives with high CO abundances, methane false positives with high CO:$CH_4$ ratios, and low noncondensable oxygen false positives where high SNR enables improved total pressure constraints. This high SNR would be demanded only for already-interesting targets, as high SNR will likely require long observation times. While not every Earth-analog biosignature and false positive could be unambiguously identified with a 0.26–1.7 μm wavelength range and SNR of 20–40—and the plausibility and contextual clues of many biosignature false positives remain an area of active research—our minimal spectral capabilities would enable a broad search for Earth-like biospheres and their false positives. Moreover, although yield modeling lies beyond the scope of this current study, the need for high SNR out to 1.7 μm for the majority of HWO targets implies that a statistically robust search for life with HWO will probably require an >8 m aperture.

### Acknowledgments

This article benefited immensely from discussions by the HWO GOMAP Working Groups, including the Living Worlds Working Group, Biosignature Interpretation sub-working group, and the Retrieval Task Group.

### Authors' Contributions

J.K.-T.: Conceptualization (equal), writing—original draft preparation (lead), and formal analysis (equal). A.G.U., M.F., and S.G.-J.: Formal analysis (colead) and writing—original draft preparation (equal). A.Y.: Conceptualization (equal), formal analysis (colead), methodology (equal), and writing—original draft preparation (equal). J.L.-Y.: Conceptualization (equal), methodology (equal), writing—original draft preparation (equal), and formal analysis (equal). T.R.: Methodology (equal) and writing—original draft preparation (supporting). S.O.: Conceptualization (equal) and writing—original draft preparation (equal). E.A., G.A., Celeste H., Chester H., N.H., É.L., A.M., M.M.M., M.N.P., B.R., E.W.S., C.S.-S., A.T., and N.W.: Conceptualization (equal) and writing—review and editing (supporting). N.L.: Conceptualization (equal), writing—original draft preparation (supporting), and writing—review and editing (supporting). S.R.: Conceptualization (supporting) and writing—review and editing (supporting).

### Data Availability

The *rfast* code used to perform the retrievals described in this study is available at https://github.com/joshuakt/rfast_for_wavelength_requirements. Posterior distributions for all biosignature and false-positive scenarios are permanently archived at https://doi.org/10.5281/zenodo.21516318.

### Author Disclosure Statement

No competing financial interests exist.

### Funding Information

This work was supported by NASA Astrophysics Decadal Survey Precursor Science grant no. 80NSSC23K1471. J.K.-T. was additionally supported by the Virtual Planetary Laboratory, a member of the NASA Nexus for Exoplanet System Science (NExSS), funded via the NASA Astrobiology Program grant no. 80NSSC23K1398, and by the Alfred P. Sloan Foundation under grant no. 2025-25204. A.Y. acknowledges support from the GSFC Sellers Exoplanet Environments Collaboration, which is supported by NASA's Planetary Science Division's Research Program. J.L.-Y. acknowledges internal support from the Johns Hopkins Applied Physics Laboratory. T.R. gratefully acknowledges support from NASA's Exoplanets Research Program (no. 80NSSC18K0349), Exobiology Program (no. 80NSSC19K0473), and Habitable Worlds Program (no. 80NSSC20K0226), as well as the Cottrell Scholar Program administered by the Research Corporation for Science Advancement. E.A.'s research was supported by an appointment to the NASA Postdoctoral Program at the NASA Goddard Space Flight Center, administered by Oak Ridge Associated Universities under contract with NASA. S.O. and E.W.S. acknowledge additional support from the NASA Interdisciplinary Consortium for Astrobiology Research with funding issued through the Alternative Earths Team (grant no. 80NSSC21K0594). S.O. additionally acknowledges support from the NASA Exobiology Program (80NSSC20K1437) and Habitable Worlds Program (80NSSC20K1409). A.M. acknowledges support from the GSFC Exoplanet Spectroscopy Technologies Work Package, which is supported by NASA's Astrophysics Science Division.

### Supplementary Material

Supplementary Data

Address correspondence to:
*Joshua Krissansen-Totton*
*Department of Earth and Space Sciences*
*University of Washington*
*1410 NE Campus Parkway*
*Seattle*
*Washington, 98195*
*USA*

*E-mail:* joshkt@uw.edu

**Abbreviations Used**

CIA = collisionally induced absorption
DTC = detection
FTT = Fischer-Tropsch-type
HWO = Habitable Worlds Observatory
IWA = Inner Working Angle
JWST = James Webb Space Telescope
LF = Land Fraction
MCMC = Markov chain Monte Carlo
NC = no constraints
NIR = near infrared
NUV = near-UV
PAL = Present Atmospheric Level
SNR = signal-to-noise
UL = upper limit
UV = ultraviolet
VIS = visible